\documentclass[onecolumn,preprint,aps,prd,epsfig] {revtex4}
\usepackage{epsfig}
\usepackage{color}
\usepackage{bm}
\usepackage{subfigure}
\usepackage{leftidx}
\usepackage{supertabular}

\newcommand{\ba}{\begin{array}}
\newcommand{\ea}{\end{array}}
\newcommand{\be}{\begin{equation}}
\newcommand{\ee}{\end{equation}}
\newcommand{\ds}{\displaystyle}
\newcommand{\veL}{\mbox{\boldmath${\rm L}$}}
\newcommand{\veS}{\mbox{\boldmath${\rm S}$}}

\begin{document}
\draft
\title{{\bf Bottomonium Spectrum with Coupled-Channel Effects }}

\author{Jia-Feng Liu$^1$ and Gui-Jun Ding$^{1, 2}$}

\affiliation{\centerline{$^1$Department of Modern
Physics,}\centerline{University of Science and Technology of China,
Hefei, Anhui 230026, China}\\
\centerline{$^2$Department of Physics, University of
Wisconsin-Madison,}\centerline{1150 University Avenue, Madison, WI
53706, USA}}

\begin{abstract}

We study the bottomonium spectrum in the nonrelativistic quark model
with the coupled-channel effects. The mass shifts and valence
$b\bar{b}$ component are evaluated to be rather large. We find that
the hadronic loop effects can be partially absorbed into a reselection of the model parameters. No bottomonium state except
$\Upsilon(5\,^1S_0)$ and $\Upsilon(5\,^3S_1)$ with mass around 10890 MeV is found in the quark models both with and without coupled-channel effects, so we suggest that
$Y_{b}(10890)$ is an exotic state beyond the quark model, if it is
confirmed to be a new resonance. The predictions for the $\chi_b(3P)$ masses are consistent with the ATLAS measurements. If some new bottomonium-like states are observed at LHCb or SuperB in the future, we can determine whether
they are conventional bottomonium or exotic states by comparing
their masses with the mass spectrum predicted in our work.

\vskip0.5cm


\end{abstract}
\maketitle
\section{introduction}

In past years, the spectroscopy of heavy flavor quarkonium has
seen great progress, particularly the charmonium spectrum. Many
charmonium-like states (such as $X(3872)$, $Y(4260)$ and so on) with
remarkable and unexpected properties have been reported. These
exotic states present great challenges to our understanding of the
structure of heavy flavor quarkonium and quantum chromodynamics (QCD) at low
energy, for a review, see
Refs.\cite{Swanson:2006st,Eichten:2007qx,Godfrey:2008nc,Brambilla:2010cs}. On the
other hand, many bottomonium states have been reported as well. In
2008, the spin-singlet pseudoscalar partner $\eta_{b}(1S)$ was found
by the Babar Collaboration with mass $M=9388.9^{+3.1}_{-2.3} ({\rm
stat}) \pm 2.7 ({\rm syst})$ MeV \cite{Aubert:2008a}. The
$\Upsilon({^3}{D}_J)$ was discovered in 2010 in the
$\pi^{+}\pi^{-}\Upsilon(1S)$ final state with mass
$M=10164.5\pm0.8({\rm stat})\pm0.5({\rm syst})$ MeV
\cite{Sanchez:2010a}. In addition, the Babar Collaboration reported the
P-wave spin-singlet $h_{b}(1P)$ via its radiative decay into
$\gamma\eta_b(1S)$ with mass $M=9902\pm4({\rm stat})\pm1({\rm
syst})$ MeV \cite{Lees:2011a}. This state is confirmed by the Belle
Collaboration \cite{Adachi:2011a}, and its mass is measured to be
$M=9898.25\pm1.06({\rm stat})^{+1.03}_{-1.07}({\rm syst})$ MeV.
Meanwhile, the radial excitation state $h_{b}(2P)$ was also found by
the Belle Collaboration with mass $M=10259.76\pm0.64({\rm
stat})^{+1.43}_{-1.03}({\rm syst})$ MeV \cite{Adachi:2011a}. Recently the ATLAS Collaboration has reported the discovery of the $\chi_b(3P)$ state through reconstruction of the radiative decay modes of $\chi_b(3P)\rightarrow\Upsilon(1S,2S)\gamma$, and its mass barycenter is measured to be $10.539\pm0.004 (\mathrm{stat.})\pm0.008 (\mathrm{syst.})$ GeV \cite{Aad:2011ih}. In
particular, the Belle Collaboration has observed an enhancement
in the production process
$e^{+}e^{-}\rightarrow\Upsilon(nS)\pi^{+}\pi^{-}(n=1,2,3)$
\cite{:2008pu}. The fit using a Breit-Wigner resonance shape yields
a peak mass of $[10888.4^{+2.7}_{-2.6}({\rm stat})\pm1.2({\rm
syst})]$ MeV and a width of $[30.7^{+8.3}_{-7.0}({\rm
stat})\pm3.1({\rm syst})]$ MeV. In the following, we shall denote
this state as $Y_b(10890)$. Moreover, the Babar Collaboration
measured the $e^{+}e^{-}\rightarrow b\bar{b}$ cross section between
10.54 GeV and 11.20 GeV \cite{:2008hx}, the $\Upsilon(10860)$ and
the $\Upsilon(11020)$ states, which are candidates of $\Upsilon(5S)$
and $\Upsilon(6S)$ respectively, were observed . Their masses and
widths are fitted to be $M_{\Upsilon(10860)}=10.876\pm0.002$ GeV,
$\Gamma_{\Upsilon(10860)}=43\pm4$ MeV,
$M_{\Upsilon(11020)}=10.996\pm0.002$ GeV and
$\Gamma_{\Upsilon(11020)}=37\pm3 $ MeV, which are different from the
previously measured values. In particular, two charged narrow
structures at 10610 MeV and 10650 MeV in the
$\pi^{\pm}\Upsilon(nS)(n=1,2,3)$ and $\pi^{\pm}h_b(mP)(m=1,2)$ have
been reported recently \cite{Collaboration:2011gja}. In summary, the current experimental data indicate that there may be
exotic bottomonium-like structures similar to the charmonium sector.
Furthermore, LHCb has begun to run \cite{LHCb}, Belle will be
updated to Belle II and a new SuperB factory will be built in Italy \cite{O'Leary:2010af}, we expect that more
heavy bottomonium states including the possible exotic extensions
will be observed in the future.

Motivated by the above exciting experimental progress in $b\bar{b}$
states, we shall carry out a careful, detailed study of bottomonium spectroscopy in this work, notably the poorly understood
higher-mass $b\bar{b}$ levels. Thus, we can determine whether future
observed bottomonium-like states could be accommodated as canonical
$b\bar{b}$ states by comparing their masses with the mass spectrum
predicted in this work. It is well-known that simple potential
models, which incorporate a color coulomb term at short distances, a
linear scalar confining term at large distances, and a
Gaussian-smeared one-gluon exchange spin-spin hyperfine
interactions, have been frequently used to describe both the charmonium
and bottomonium spectrums. Generally, the mixture between the quark
model $b\bar{b}$ basis states and the two-meson continuum has been neglected
in these models, which are called ``quenched" quark models.

The effects of the ``unquenched quark model" including virtual
hadronic loops have been studied extensively in the framework of the
coupled-channel method
\cite{Tornqvist:1979hx,Ono:1983rd,Tornqvist:1995kr,vanBeveren:1979bd,Eichten:1974af}.
The hadronic loop has turned out to be highly non-trivial, it can
give rise to mass shifts to the bare hadron states and contribute
continuum components to the physical hadron states. The possibility
that loop effects may be responsible for the anomalously low masses
of the new narrow charm-strange states $D^{*}_{s0}(2317)$ and
$D_{s1}(2460)$ has been suggested by several groups
\cite{Barnes:2003dj,vanBeveren:2003kd,Hwang:2004cd,Simonov:2004ar}.
The hadronic loop in charmonium has been explored as well, and the mass
shifts and continuum mixing due to loops of $D$, $D^{*}$, $D_s$ and
$D^{*}_s$ meson pairs have been studied extensively
\cite{Eichten:2004uh,Kalashnikova:2005ui,Pennington:2007xr,Barnes:2007xu,Danilkin:2009hr,Zhang:2009bv,Li:2009ad,Ortega:2010qq}.
Both the mass shifts and the two-meson continuum components of the
physical charmonium states were found to be rather large. In
particular, a $J^{PC}=1^{++}$ state with mass about 3872 MeV could
possibly be generated dynamically.

Inspired by the large physical effects of hadronic loops in both the
$D_{sJ}$ and charmonium states, we expect that the virtual hadronic
loop should also play an important role in bottomonium spectroscopy. In this work, we shall study the bottomonium spectrum in
detail, and take the hadronic loop effects into account. This
paper is organized as follows. We present the framework of the
coupled-channel analysis in section II. The non-relativistic
potential model is outlined in section III. Section IV is devoted to
the numerical results for the masses of the bottomonium states with
and without hadronic loop effects, as well as phenomenological implications.
We present our conclusions and discussion in section V.

\section{Formalism of coupled-channel analysis}

\begin{figure}[hptb]
\begin{center}
\includegraphics[width=0.50\textwidth]{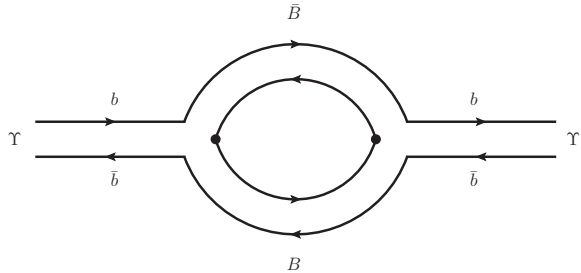}
\caption{\label{figcc} Coupling of $b\overline{b}$ states to the
$B\bar{B}$ mesons loop.}
\end{center}
\end{figure}

In bottomonium, the process
$(b\bar{b})\rightarrow(b\bar{n})(n\bar{b})$ via light quark pair $n\bar{n}$ creation would induce the hadronic loop shown
in Fig.\ref{figcc}, where the initial bottomonium decays into
intermediate virtual $B\bar{B}$ states and then reforms the original
bottomonium state. Here $B(\bar{B})$ denotes a general $B(\bar{B})$
meson, it can be $B(\bar{B})$, $B_s(\bar{B}_s)$,
$B^{*}(\bar{B}^{*})$ or $B^{*}_s(\bar{B}^{*}_s)$, the same
convention will be used henceforth without speci?cation.
Since the open-flavor decay couplings of bottomonium states to
two-body $B\bar{B}$ final states are large, the resulting loop
effects should be important. This kind of virtual hadronic loop is
universal, but it is not usually included in quark potential
models and is only partially present in the quenched lattice QCD.
The coupled-channel model is an appropriate framework for analyzing these
hadronic loop effects
\cite{Tornqvist:1979hx,Ono:1983rd,Tornqvist:1995kr,vanBeveren:1979bd,Eichten:1974af}.
In the simplest version of the coupled-channel model
\cite{Kalashnikova:2005ui}, the full hadronic state is represented
as
\begin{equation}
|\Psi \rangle = \left(\sum_{\alpha}c_{\alpha}|\psi
_{\alpha}\rangle\atop \sum_{B\bar{B}} \chi_{B\bar{B}}(\mathbf{p})
|\phi_{B}\phi_{\bar{B}}\rangle \right), \label{state}
\end{equation}
with the normalization condition
$\sum_{\alpha}|c_{\alpha}|^2+\sum_{B\bar{B}}\int
d^3\mathbf{p}|\chi_{B\bar{B}}(\mathbf{p})|^2=1$.
$|\psi_{\alpha}\rangle$ denotes the bare confined $b\bar{b}$ states
with the probability amplitude $c_{\alpha}$, $\phi_{B}$
($\phi_{\bar{B}}$) is the $\bar{b}n$($b\bar{n}$) eigenstate describing
the $B$($\bar{B}$) meson, and $\chi_{B\bar{B}}(\mathbf{p})$ is the
wavefunction in the two-meson channel
$|\phi_{B}\phi_{\bar{B}}\rangle$. The wavefunction $|\Psi\rangle$
obeys the equation
\begin{equation}
\label{cce}{\mathcal H} |\Psi\rangle = M |\Psi\rangle,\quad
\mathcal{H}= \left(\begin{array}{cc}
H_0&H_I\\
H_I&H_{B\bar{B}}
\end{array}
\right), \label{ham}
\end{equation}
where $H_0$ is the Hamiltonian for the valence $b\bar{b}$ system,
with the eigenstates determined by $H_0 |\psi_{\alpha}\rangle =
M_{\alpha} |\psi_{\alpha}\rangle$. The Hamiltonian $H_{B\bar{B}}$
acts between the constituents of $B$ and $\bar{B}$ separately,
where the interactions between $B$ and $\bar{B}$ are neglected. The
continuum two-meson state $|\phi_{B}\phi_{\bar{B}}\rangle$ is the
eigenstate of $H_{B\bar{B}}$,
\begin{eqnarray}
\label{radd1}
H_{B\bar{B}}|\phi_{B}\phi_{\bar{B}}\rangle&=&(E_{B}+E_{\bar{B}})|\phi_{B}\phi_{\bar{B}}\rangle
\simeq\Big(m_{B}+m_{\bar{B}}+\frac{\mathbf{p}^2}{2\mu_{B\bar{B}}}\Big)|\phi_{M_1}\phi_{M_2}\rangle
\end{eqnarray}
where $E_{B}=\sqrt{m^2_{B}+\mathbf{p}^2}$,
$E_{\bar{B}}=\sqrt{m^2_{\bar{B}}+\mathbf{p}^2}$, $m_{B}$ and
$m_{\bar{B}}$ are the masses of $B$ and $\bar{B}$ respectively, and
$\mu_{B\bar{B}}=\frac{m_{B}m_{\bar{B}}}{m_{B}+m_{\bar{B}}}$ is the
reduced mass of the two-meson system. $H_I$ couples the bare state
$|\psi_{\alpha}\rangle$ with the two-body continuum
$|\phi_{B}\phi_{\bar{B}}\rangle$. Let us consider one bare state
$|\psi_0\rangle$, the matrix element of $H_I$ is of the following
form:
\begin{equation}
\label{radd2}\langle
\phi_{B}\phi_{\bar{B}}|H_I|\psi_0\rangle=h^{0}_{B\bar{B}}(\mathbf{p})
\end{equation}
Substituting Eq.(\ref{radd1}) and Eq.(\ref{radd2}) into Eq.(\ref{cce}), we get the system of coupled equations for $c_0$ and
$\chi_{B\bar{B}}(\mathbf{p})$,
\begin{equation} \left\{
\begin{array}{c}
\ds c_0M_0 + \sum_{B\bar{B}} \int h^{0}_{B\bar{B}}(\mathbf{p})\chi_{B\bar{B}}(\mathbf{p})d^3\mathbf{p} = Mc_0\\
(E_{B}+E_{\bar{B}})\chi_{B\bar{B}}(\mathbf{p}) +
c_0h^{0}_{B\bar{B}}(\mathbf{p}) = M \chi_{B\bar{B}}(\mathbf{p})
\end{array}
\right. \label{system}
\end{equation}
This coupled-channel equation can be solved straightforwardly, and we finally obtain the master equation
\begin{equation}
\label{radd3}M-M_0+\sum_{B\bar{B}}\Pi_{B\bar{B}}(M)=0.
\end{equation}
Here $\Pi_{B\bar{B}}(M)$ is the self-energy function for the
hadronic loop induced by the intermediate states $B$ and $\bar{B}$,
it is explicitly given by
\begin{equation}
\label{radd4}\Pi_{B\bar{B}}(M)=\int
\frac{|h^{0}_{B\bar{B}}(\mathbf{p})|^{2}}{E_{B}+E_{\bar{B}}-M-i\epsilon}\,
d^3\mathbf{p}
\end{equation}
Using the relation between the helicity amplitude $h^{0}_{B\bar{B}}$
and the partial wave amplitude ${\cal M}_{LS}$ \cite{Jacob:1959at},
we have
\begin{eqnarray}
\nonumber&&\Pi_{B\bar{B}}(M)=\int \limits_{0}^{\infty}dp
\frac{p^{2}}{E_B+E_{\bar{B}}-M-i\epsilon}\int
d\Omega_{\hat{\mathbf{p}}} |h^{0}_{B\bar{B}}(\mathbf{p})|^{2}\\
\nonumber\,~&&=\int \limits_{0}^{\infty}dp
\frac{p^{2}}{E_B+E_{\bar{B}}-M-i\epsilon}
\sum\limits_{LS} |{\cal M}_{LS}|^{2}\\
\,~&&={\cal P}\int \limits_{0}^{\infty}dp
\frac{p^{2}}{E_B+E_{\bar{B}}-M} \sum\limits_{LS} |{\cal
M}_{LS}|^{2}+i \pi\Big(\frac{pE_{B}E_{\bar{B}}}{M}\sum\limits_{LS}
|{\cal M}_{LS}|^{2}\Big)\Big|_{E_B+E_{\bar{B}}=M} \label{t1}
\end{eqnarray}
Note that the imaginary part of the self-energy $\Pi_{B\bar{B}}(M)$ arises only if the initial hadron mass is above the intermediate $B\bar{B}$ threshold. Comparing with the two-body strong decay width shown in
Eq.(\ref{add2}), it is obvious that the imaginary part is exactly equal
to half of the decay width, if the decay is not forbidden
kinematically. For the bottomonium state above the threshold, its mass is determined by the real part of the master equation Eq.(\ref{radd3}) \cite{Barnes:2007xu,Li:2009ad,Pennington:2007xr}. The squared absolute value $|c_0|^2$ is proportional
to the probability that the physical energy eigenstate is in the
$b\bar{b}$ configuration, and the $b\bar{b}$ component is given by
\begin{equation}
P_{b\bar{b}}=1\Big/\Big(1+\sum_{B\bar{B}}\int\frac{|h^{0}_{B\bar{B}}(\mathbf{p})|^2}{(E_B+E_{\bar{B}}-M)^2}\,d^{3}\mathbf{p}\Big)
\end{equation}
In this work the effects of virtual hadronic loops will be considered
in the above framework, and we shall sum over the contributions of
intermediate loops from two stable S-wave mesons $B\bar{B}$,
$B\bar{B}^{*}$, $B^{*}\bar{B}$, $B^{*}\bar{B}^{*}$, $B_s\bar{B}_s$,
$B_s\bar{B}^{*}_s$, $B^{*}_s\bar{B}_s$ and $B^{*}_s\bar{B}^{*}_s$,
where antiparticles are indicated explicitly .

\section{Non-relativistic potential model }

We use the standard non-relativistic potential model to describe the
bare valence $b\bar{b}$ states. Its Hamiltonian is of the form
\begin{eqnarray}
\nonumber &&H_0=\frac{\mathbf{p}^2}{m_b}+V_{cou}(r)+V_{con}(r)+C+V_{sd}(r)\\
&&{V}_{cou}(r)=- \frac{4}{3} \frac{\alpha_s}{r}, \quad
{V}_{con}(r)=\sigma r, \label{potential}
\end{eqnarray}
where $m_b$ is the bottom quark mass, $V_{cou}(r)$ is the well-known
color-Coulomb force, and $V_{con}(r)$ denotes the linear confinement
potential. To restore the hyperfine and fine structure of the
bottom spectrum, one needs to introduce the Fermi-Breit relativistic
corrections term $V_{sd}(r)$, which includes spin-spin, spin-orbit and
tensor force. This is explicitly given by
\begin{equation}
V_{sd}(r)=V_{SS}(r)+V_{LS}(r)+V_{T}(r), \label{SD}
\end{equation}
where $V_{SS}(r)$, $V_{LS}(r)$ and $V_{T}(r)$ are spin-spin,
spin-orbit and tensor operators respectively, and $V_{SS}$ is the
contact hyperfine interaction,
\begin{equation} V_{SS}(r)=\frac{2\veS_{b} \cdot
\veS_{\bar
b}}{3m_b^2}[\Delta({V}_{cou}(r))]=\frac{32\pi\alpha_s}{9m_b^2}~{\tilde
\delta}(r)~\veS_{b} \cdot \veS_{\bar b}, \label{HF}
\end{equation}
where $\veS_{b}$ is the spin of the bottom quark and $\veS_{\bar b}$ is
the spin of the anti-bottom quark. The Gaussian smearing of the
hyperfine interaction is introduced here,
\begin{equation} {\tilde
\delta}(r)=(\frac{\kappa}{\sqrt{\pi}})^{3}e^{-\kappa^{2}r^{2}}
\end{equation}
The spin-orbit term is given by
\begin{equation}
V_{LS}(r)=[3\frac{d}{dr}{V}_{cou}(r)-\frac{d}{dr}{V}_{con}(r)]\frac{\veL
\cdot
\veS}{2m_b^2r}=(\frac{4\alpha_s}{r^{3}}-\frac{\sigma}{r})\frac{\veL
\cdot \veS}{2m_b^2},
\end{equation}
where $\veS=\mathbf{S}_b+\mathbf{S}_{\bar{b}}$ is the total spin, and
$\veL$ is the relative angular momentum between $b$ and $\bar{b}$.
Finally, the tensor term is
\begin{equation}
V_{T}(r)=\frac{T_{b\overline{b}}}{m_b^2}[\frac{1}{r}\frac{d}{dr}{V}_{cou}(r)-\frac{d^{2}}{dr^{2}}{V}_{cou}(r)]=\frac{4\alpha_{s}}{m_b^2r^3}T_{b\overline{b}}
\end{equation}
where $T_{b\overline{b}}$ is the well-known tensor force
operator,
\begin{equation}
T_{b\overline{b}}=3(\mathbf{S}_{b}\cdot
\hat{\mathbf{r}})(\mathbf{S}_{\bar{b}}\cdot
\hat{\mathbf{r}})-\mathbf{S}_{b}\cdot \mathbf{S}_{\bar{b}}
\end{equation}
The spin-dependent term $V_{sd}$ is relativistically suppressed
with respect to $V_{cou}$ and $V_{con}$, it is generally treated using
leading order perturbation theory.

\section{Predictions for the bottomonium spectrum}
The interaction Hamiltonian $H_I$, which couples the bare valence
$b\bar{b}$ with the two-body $B\bar{B}$ continuum, is an essential
element of this formalism. In this work, we shall use the well-established $^3P_0$ model
\cite{Micu:1969a,orsay,Geiger:1994kr,Ackleh:1996yt} to describe the
mixing between the bare $b\bar{b}$ state and the open bottom meson
pair $b\bar{q}$ and $q\bar{b}$. The ${^3}{P}{_0}$ model assumes that the Okubo, Zweig, and Iizuka (OZI) rule-allowed
strong decay takes place via the creation of a quark-antiquark pair with $J^{PC}=0^{++}$ from the vacuum. The
$q\bar{q}$ pair production is described by the Hamiltonian,
\begin{equation}
\label{1}H_I=g\sum_q\int
d^3\mathbf{x}\;\overline{\psi}_q(x)\;\psi_q(x)
\end{equation}
where $\psi_q$ is the Dirac quark field. Following the conventional
calculating method in $^3P_0$ model, one can then straightforwardly
evaluate the valence-continuum coupling matrix element
$h^{0}_{B\bar{B}}$ in Eq.(\ref{radd2}). Here we will use simple
harmonic oscillator (SHO) wavefunctions for the involved mesons,
with a universal oscillator parameter $\beta$. The SHO wavefunction
enables analytical calculation of the transition amplitudes, and it
turns out to be a good approximation. In Appendix A, we present the
analytical $^3P_0$ amplitudes ${\cal M}_{LS}$ for two S-wave
final state channels. Since the oscillator parameters of the initial and final states are different in these analytical expressions, for an initial state of higher radial excited bottomonium, we can obtain the required transition amplitudes by simply taking derivatives, as is shown in Appendix B.
Here we will take the typical values $\beta=0.4$ GeV for numerical
calculations, this value was frequently adopted in the literature
\cite{Ackleh:1996yt,Barnes:1996ff,Barnes:2002mu,Liu:2010a}, and it turned out to be a reasonable zeroth-order approximation in coupled-channel calculations as well \cite{Barnes:2007xu}.  The
parameter $g$ has the form $g=2m_{q}\gamma$, where $m_q$ is the
constituent quark mass, and $\gamma$ is the effective strength of
pair creation. While for the creation of strange quarks, the
effective strength $\gamma_{s}$=$(m_{q}/m_{s})\gamma$ is used,
following Ref. \cite{Kalashnikova:2005ui}, we shall take
$\gamma$=0.322. The masses of constituent quarks are chosen to be
$m_u=m_d=0.33$ GeV and $m_s=0.55$ GeV as usual. In the following, we
first present the predictions for the $b\bar{b}$ spectrum in the
nonrelativistic potential model of section III, then we include the
coupled-channel effects, and the phenomenological implications are
discussed.
\subsection{Bottomonium spectrum in conventional non-relativistic potential model}

As usual, we can determine the bare $b\bar{b}$ mass spectrum by
solving the following Schr$\ddot{\rm o}$dinger equation numerically,
\begin{equation}
-\frac{1}{m_b}\frac{d^{2}}{dr^{2}}u_{nl}(r)+[V(r)+\frac{l(l+1)}{m_b
r^{2}}]u_{nl}(r)=E_{nl}u_{nl}(r)
\end{equation}
where $E_{nl}$ is the energy eigenvalue, the potential $V(r)$ is the
leading order one with $V(r)=V_{cou}(r)+V_{con}(r)+C$. The
wavefunction $\Psi_{nlm}$ of the system is closely related to
$u_{nl}(r)$ by
\begin{equation}
\Psi_{nlm}(r,\theta,\phi)=R_{nl}(r)Y_{lm}(\theta,\varphi) \quad {\rm
and} \quad u_{nl}(r)=rR_{nl}(r)
\end{equation}
The mass splitting within the multiplets is determined by the
spin-dependent term $V_{sd}(r)$, which is taken to be a
perturbation. As a result, we need to calculate the expectation
value of $V_{sd}(r)$ between the leading order wavefunction
$\Psi_{nlm}$, in order to get the bottomonium mass fine splitting.
The only term which gives a nonvanishing contribution to
$n\,{^1}{S}_{0}-n\,{^3}{S}_{1}$ fine splitting is the spin-spin
term.
Since the expectation value of $\veS_{b} \cdot \veS_{\bar b}$ in the
${^1}{S}{_0}$ and ${^3}{S}{_1}$ states is $-3/4$ and $1/4$
respectively, the bare masses of S-wave states are
\begin{eqnarray}
&&M_{0}(n{^1}{S}{_0})=E_{n0}-\frac{3}{4}\frac{32\pi\alpha_{s}}{9{m^2_{b}}}\langle\tilde{\delta}(r)\rangle\\
&&M_{0}(n{^3}{S}{_1})=E_{n0}+\frac{1}{4}\frac{32\pi\alpha_{s}}{9{m^2_{b}}}\langle\tilde{\delta}(r)\rangle
\end{eqnarray}
where
\begin{equation}
\label{exp}\langle\tilde{\delta}(r)\rangle=(\frac{\kappa}{\sqrt{\pi}})^{3}\int_{0}^{\infty}e^{-\kappa^{2}r^{2}}u^2_{n0}(r)dr
\end{equation}
For the $n\,{^1}{P}{_1}-n\,{^3}{P}{_J}$,
$n\,{^1}{D}_{2}-n\,{^3}{D}_{J}$, $n\,{^1}{F}_{3}-n\,{^3}{F}_{J}$
cases, the spin-orbital and tensor terms contribute as well. The
corresponding bare masses are given by
\begin{eqnarray}
&&M_{0}(n\,{^1}{P}_{1})=E_{n1}-\frac{3}{4}\frac{32\pi\alpha_{s}}{9{m^2_{b}}}\langle\tilde{\delta}(r)\rangle\\
&&M_{0}(n\,{^3}{P}_{J})=E_{n1}+\frac{1}{4}\frac{32\pi\alpha_{s}}{9{m^2_{b}}}\langle\tilde{\delta}(r)\rangle+\frac{1}{{m^2_{b}}}\Big(A_{J}\alpha_{s}\langle\frac{1}{r^3}\rangle+B_{J}\sigma\langle\frac{1}{r}\rangle\Big)\\
&&M_{0}(n\,{^1}{D}_{2})=E_{n2}-\frac{3}{4}\frac{32\pi\alpha_{s}}{9{m^2_{b}}}\langle\tilde{\delta}(r)\rangle\\
&&M_{0}(n\,{^3}{D}_{J})=E_{n2}+\frac{1}{4}\frac{32\pi\alpha_{s}}{9{m^2_{b}}}\langle\tilde{\delta}(r)\rangle+\frac{1}{{m^2_{b}}}\Big(C_{J}\alpha_{s}\langle\frac{1}{r^3}\rangle+D_{J}\sigma\langle\frac{1}{r}\rangle\Big)\\
&&M_{0}(n\,{^1}{F}_{3})=E_{n3}-\frac{3}{4}\frac{32\pi\alpha_{s}}{9{m^2_{b}}}\langle\tilde{\delta}(r)\rangle\\
&&M_{0}(n\,{^3}{F}_{J})=E_{n3}+\frac{1}{4}\frac{32\pi\alpha_{s}}{9{m^2_{b}}}\langle\tilde{\delta}(r)\rangle+\frac{1}{{m^2_{b}}}\Big(E_{J}\alpha_{s}\langle\frac{1}{r^3}\rangle+F_{J}\sigma\langle\frac{1}{r}\rangle\Big)\\
&&M_{0}(n\,{^1}{G}_{4})=E_{n4}-\frac{3}{4}\frac{32\pi\alpha_{s}}{9{m^2_{b}}}\langle\tilde{\delta}(r)\rangle\\
&&M_{0}(n\,{^3}{G}_{J})=E_{n4}+\frac{1}{4}\frac{32\pi\alpha_{s}}{9{m^2_{b}}}\langle\tilde{\delta}(r)\rangle+\frac{1}{{m^2_{b}}}\Big(G_{J}\alpha_{s}\langle\frac{1}{r^3}\rangle+H_{J}\sigma\langle\frac{1}{r}\rangle\Big)
\end{eqnarray}
Here $\langle\ldots\rangle$ denotes the expectation value, which is defined
in the same way as that in Eq.(\ref{exp}). The coefficients $A_{J}$,
$B_{J}$, $C_{J}$, $D_{J}$, $E_{J}$, $F_{J}$, $G_J$ and $H_J$ for
each case are listed in Table \ref{tab:coefficients}. Note that the tensor force $T_{b\bar{b}}$ could lead to the so-called S$-$D mixing. However, the mass splitting induced by the S$-$D mixing is a second-order perturbation effect of the hyperfine interactions $V_{sd}$, thus its contribution is rather small and hence is neglected here. We find that
the numerical value $\langle\tilde{\delta}(r)\rangle$ is extremely
small except for the $S-$wave states, since it would be proportional
to the zero point value of the wavefunction, i.e., it is exactly
zero if we don't smear the hyperfine interaction. As a result, the
following sum rules are satisfied quite well,
\begin{eqnarray}
\label{eq:sum1}&&M_{0}(n\,{^1}{P}_{1})\simeq(5M_{0}(n\,{^3}{P}_{2})+3M_{0}(n\,{^3}{P}_{1})+M_{0}(n\,{^3}{P}_{0}))/9\\
\label{eq:sum2}&&M_{0}(n\,{^1}{D}_{2})\simeq(7M_{0}(n\,{^3}{D}_{3})+5M_{0}(n\,{^3}{D}_{2})+3M_{0}(n\,{^3}{D}_{1}))/15\\
\label{eq:sum3}&&M_{0}(n\,{^1}{F}_{3})\simeq(9M_{0}(n\,{^3}{F}_{4})+7M_{0}(n\,{^3}{F}_{3})+5M_{0}(n\,{^3}{F}_{2}))/21\\
\label{eq:sum4}&&M_{0}(n\,{^1}{G}_{4})\simeq(11M_{0}(n\,{^3}{G}_{5})+9M_{0}(n\,{^3}{G}_{4})+7M_{0}(n\,{^3}{G}_{3}))/27
\end{eqnarray}

\begin{table}[t]
\renewcommand{\arraystretch}{0.5}
\begin{tabular}
{|c|c|c|c|c|c|c|c|c|} \hline\hline
~J~ &$A_{J}$&$B_{J}$&$C_{J}$&$D_{J}$&$E_{J}$&$F_{J}$&$G_{J}$&$H_{J}$\\
\hline
$0$&$-16/3$&1&\quad&\quad&\quad&\quad&\quad&\quad\\
$1$&$-4/3$&1/2&$-20/3$&3/2&\quad&\quad&\quad&\quad\\
$2$&28/15&$-1/2$&$-4/3$&1/2&$-128/15$&2&\quad&\quad\\
$3$&\quad&\quad&80/21&$-1$&$-4/3$&1/2&$-220/21$&5/2\\
$4$\quad&\quad&\quad&\quad&&52/9&$-3/2$&$-4/3$&1/2\\
$5$\quad&\quad&\quad&\quad&\quad&\quad&\quad&256/33&$-2$\\
\hline\hline
\end{tabular}
\renewcommand{\arraystretch}{0.5}
\caption{\label{tab:coefficients} Numerical values for the $A_{J}$,
$B_{J}$,$C_{J}$, $D_{J}$, $E_{J}$, $F_{J}$, $G_{J}$ and $H_{J}$
coefficients.}
\end{table}

For experimental input we use the masses of the 14 reasonably well-established $b\bar{b}$ states, which are given in Table
\ref{tab:allpredicted}. Here we don't include the $\Upsilon(5S)$ and
$\Upsilon(6S)$ candidates $\Upsilon(10860)$ and $\Upsilon(11020)$,
since their masses measured by the Babar Collaboration are different
from the Particle Data Group (PDG) averages \cite{pdg}. The parameters that follow from
fitting these masses are $\alpha_{s}$=0.3840, $\sigma$=0.9155$\mathrm{
GeV/fm}$, $C=-0.7825$ GeV, $m_{b}$=5.19 GeV and $\kappa$=2.3 GeV. In the classical Godfrey-Isgur quark model \cite{Godfrey:1985xj}, the effective strong coupling constant $\alpha_s$ at the bottomonium scale and the string tension $\sigma$ are determined to be about 0.25 and 0.18 ${\rm GeV}^2$ respectively. Clearly the fitting value for $\sigma$ is approximately the same as that in \cite{Godfrey:1985xj}, while $\alpha_s$ is found to be somewhat larger than that of \cite{Godfrey:1985xj}. Note that $\alpha_{s}$ is dissociated from the effective strong coupling constant determined by the hadronic width of the quarkonium in conventional potential models, it is only a purely phenomenological strength parameter of the short-range potential \cite{Eichten:1974af}. Given these values, we can predict the masses of the currently
unknown $b\bar{b}$ states. The predicted spectrum is shown in Table
\ref{tab:allpredicted}, and we see that the mass sum rules in Eq.
(\ref{eq:sum1})-Eq.(\ref{eq:sum4}) are really satisfied rather well.

\subsection{Bottomonium spectrum with coupled-channel effects}
\begin{figure}[hptb]
\begin{center}
\includegraphics[width=0.75\textwidth]{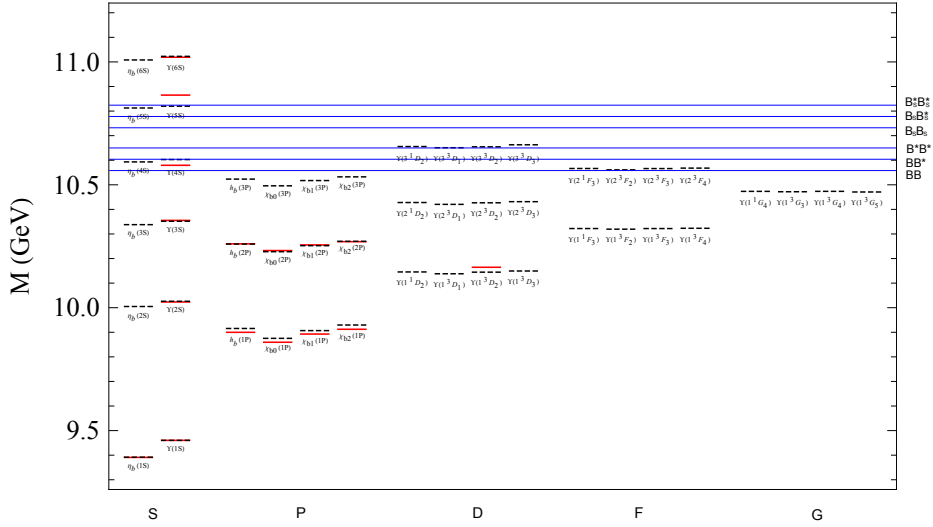}
\end{center}
\caption{\label{figbb} Predicted and observed spectrums of
bottomonium states. The solid lines are experimentally measured data,
and the dashed lines are predicted masses with coupled-channel effects
included. Various $B\bar{B}$ thresholds are also shown.}
\end{figure}

Following the formalism presented in section II, we shall take the coupled-channel effects into
account. By performing a fit to the 14
established experimental states given in Table
\ref{tab:allpredicted}, the best values of the parameters are
determined to be $\alpha_s$=0.418, $\sigma$=0.818$\mathrm{GeV/fm}$,
$C=-0.62376$ GeV, $m_b$=5.18 GeV and $\kappa$=2.85 GeV. Note that the value of $\alpha_s$ here is larger than that in Ref.\cite{Godfrey:1985xj}, while $\sigma$ is smaller than the fitting value of \cite{Godfrey:1985xj}. We can now
straightforwardly evaluate the bare state masses, the mass shifts
due to $B\bar{B}$ loops and the predictions for the bottomonium
masses with coupled-channel effects included. The results are given
in Table \ref{tab:shift1} and Table \ref{tab:shift2}, where the
$b\bar{b}$ component is presented as well. It can be seen that
the mass sum rules in Eq. (\ref{eq:sum1})-Eq.(\ref{eq:sum4}) remain approximately
intact. In order to clearly see the predictions for
the bottomonium spectrum, we further plot the predicted masses in
Fig. \ref{figbb}.

From Table \ref{tab:shift1} and Table \ref{tab:shift2}, we can see
that the mass shifts are predicted to be close to each other, they
are of order 100 MeV, although individual $B\bar{B}$ loops make
different contributions to the mass shift of each state. Moreover,
we see that the total $B\bar{B}$ components are rather large, they
mostly scatter in the range of $0.1\sim0.25$. We conclude that the
results for the hadronic loop contributions to the bottomonium
states are consistent with the general loop theorem derived in Ref.
\cite{Barnes:2007xu}. We note that the mass shifts in the charmonium
sector are predicted to be around 200 MeV \cite{Kalashnikova:2005ui}
(a larger value of about 500 MeV was suggested in \cite{Barnes:2007xu}),
and the two-meson continuum components can be as large as 0.5.
Therefore the hadronic loop effects in bottomonium are much smaller
than those in the charmonium sector.

In order to compare the mass spectrum in the non-relativistic
potential model, we show the predicted mass with coupled-channel
effects in Table \ref{tab:allpredicted} as well. For the 16 observed
states, both predictions agree well with the measured masses
except $\Upsilon(4\,^3S_1)$, $\Upsilon(5\,^3S_1)$ and
$\Upsilon(6\,^3S_1)$. We see that the masses of $\Upsilon(4\,^3S_1)$
and $\Upsilon(6\,^3S_1)$ are predicted to be rather close to the
observation after the coupled-channel effects are included, but the
mass of $\Upsilon(5\,^3S_1)$ is smaller than its measured value
about 51 MeV. Whereas the non-relativistic potential model without
hadronic loop predicts the $\Upsilon(5\,^3S_1)$ mass successfully, the
departures of predictions from observation are 41.6 MeV and 57.4 MeV
respectively for the $\Upsilon(4\,^3S_1)$ and $\Upsilon(6\,^3S_1)$
states. The current experimental data indicates that coupled-channel
effects can improve the agreement with observations, although it is
not as satisfactory for the $\Upsilon(5\,^3S_1)$ state. But this is not
the whole story, because the recently measured masses of
$\Upsilon(5\,^3S_1)$ and $\Upsilon(6\,^3S_1)$ by the Babar
Collaboration are different from the world averages. More precise
measurements of the $\Upsilon(5\,^3S_1)$ and $\Upsilon(6\,^3S_1)$ states
would be helpful in understanding the unquenched effects induced by
hadronic loops in the bottomonium spectrum. Regarding the mass barycenter of the $\chi_b(3P)$ states, it is predicted to be 10523.3 MeV and 10537.9 MeV respectively for the included and not included coupled-channel effects. Both predictions are consistent with the ATLAS measurement, although the former is slightly smaller.

Inspecting the mass predictions shown in Table
\ref{tab:allpredicted}, we can't find a state with mass around 10890
MeV except $\Upsilon(5\,^1S_0)$ and $\Upsilon(5\,^3S_1)$, even if coupled-channel effects are
considered. Therefore if $Y_b(10890)$ is confirmed to be a new
resonance by future experiments, it should not be a canonical
bottomonium state. Many interpretations have been proposed so far. A
possible explanation is the existence of a tetraquark state
$[bq][\bar{b}\bar{q}]$ \cite{Karliner:2008rc,Ali:2010pq}. Another
explanation is a $b\bar{b}$ counterpart to the $Y(4260)$ state, which may
overlap with $\Upsilon(5S)$ \cite{Hou:2006it}. Other interpretations
such as final state interactions and so forth have been suggested as well
\cite{Meng:2008dd}. In the same manner, if some new bottomonium-like
state is observed by LHCb or SuperB in the future, we can determine
whether the state could be accommodated as a conventional quark
model state by comparing its mass with our predictions, and we can determine its assignment if it is.

Mixing between two bare $b\bar{b}$ states could arise through the hadron loop, provided that both bottomonium states could couple to the same intermediate $B\bar{B}$ state. Similar to the S$-$D mixing, this kind of mixing would introduce corrections to the mass of the physical bottomonium. However, its contributions are of higher order in the valence-continuum coupling Hamiltonian $H_I$ with respect to the loop contributions discussed above. Moreover, as is stated by the loop theorem of Ref. \cite{Barnes:2007xu}, if the mass difference between various intermediate loop mesons is neglected, the mixing amplitude between two valence $b\bar{b}$ states vanishes unless both the orbital angular momentum and the spin of the two states are the same. Concrete numerical calculations show that this mixing effect is really quite small \cite{Barnes:2007xu}, and the same results are found in Ref. \cite{Kalashnikova:2005ui}. As a result, we have not considered this effect in the present work.

Finally, we note that although the bottomonium mass spectrums
predicted in models with and without coupled-channel effects are not
drastically different from each other, i.e., part of the hadronic
loop effects can be absorbed into the redefinition of the model
parameters, the underlying physics is different. In the scenario
with coupled-channel effects, there are sizable $B\bar{B}$
components in the physical bottomonium states. As a result, the
production and decay of bottomonium would be different from those
predicted in the quenched quark model. The
coupled-channel effects in hadronic transitions of bottomonium have been studied in Ref. \cite{Zhou:1990ik}, and it was found that the inclusion of the coupled-channel effects can really improve the theory of hadronic transitions. The mixture of $B\bar{B}$
continuum may also be important in understanding some anomalous
observations, since the same has turned out to be true in the
charmonium sector \cite{Ono:1983rd,Eichten:1974af,Guo:2010ak,Badalian:2012jz}. This topic
deserves much laborious and complex work and is beyond the scope of the present work.

\section{Conclusions and discussions }

Motivated by the recent experimental progress on the bottomonium
spectrum,, we speculate that some bottomonium-like states may be observed
in the future similar to the charmonium sector, particularly with the
running of LHCb and SuperB. Because the coupled-channel effects are
very important in understanding the nature of the newly observed
charmonium-like states, such as $X(3872)$, the same is expected to
be true in the bottomonium sector. In this work, we investigate the
hadronic loops effects in the bottomonium spectrum. The coupling between
the valence $b\bar{b}$ states and the two-meson $B\bar{B}$ continuum
is described in terms of $^3P_0$ model. The mass shifts and the
$b\bar{b}$ component of the physical bottomonium are calculated in
detail. We find that the mass shifts for all states are similar to
each other, they are around 100 MeV, although the contributions of
individual loops are different for each state. The two-meson
continuum $B\bar{B}$ components are found to be rather large as
well. The hadronic loop effects in bottomonium turn out to be
smaller than the ones in the charmonium sector. Moreover, we
evaluate the mass spectrum in the conventional constituent quark model,
where coupled-channel effects are not taken into account. We
find that the hadronic loop effects can be partially absorbed into
a reselection of the model parameters. Since the potential models both
with and without coupled-channel effects don't predict a state with
mass around 10890 MeV except for $\Upsilon(5\,^1S_0)$ and $\Upsilon(5\,^3S_1)$, we conclude that
$Y_b(10890)$ should be an exotic state beyond the quark model, if it is
confirmed to be a new resonance. Our prediction for the mass of $\chi_{b}(3P)$ states is consistent with the recent ATLAS measurement. Moreover, with the mass spectrum
predicted in our work, we can determine
whether a new bottomonium-like state observed in the future can be
accommodated as a canonical bottomonium, and we can determine its assignment if it is.

\begin{acknowledgments}
We are grateful to Professor Mu-Lin Yan and Professor Dao-Neng Gao for
stimulating discussions. Jia-Feng Liu would like to express special
thanks to Hao-Ran Chang for beneficial suggestions and to Song-Bin
Zhang for useful help with the numerical calculation. This work is
supported by the National Natural Science Foundation of China under
Grant No.10905053, Chinese Academy KJCX2-YW-N29 and the 973 project
with Grant No. 2009CB825200. Jia-Feng Liu is supported in part by
the National Natural Science Foundation of China under Grant
No.10775124, No.11075149 and No.10975128.
\end{acknowledgments}


\newpage

\begin{center}
{\bf Appendix A: $^3P_0$ transition amplitudes}
\end{center}
\setcounter{equation}{00}
\renewcommand{\theequation}{A.\arabic{equation}}

The derivation of the $^3P_0$ matrix elements has been discussed in
detail in Ref.
\cite{Ackleh:1996yt,Barnes:1996ff,Barnes:2002mu,Liu:2010a}. Starting
from the pair production Hamiltonian $H_I$ given in Eq.(\ref{1}),
one can straightforwardly evaluate the $H_I$ matrix element $h_{fi}$
for the transition $A(b\bar{b})\rightarrow B(q\bar{b})+C(b\bar{q})$
in terms of overlap integrals in flavor, spin and spatial spaces,
\begin{eqnarray}
\nonumber h_{fi}&&=\langle BC|H_I|A\rangle_{a}+\langle BC|H_I|A\rangle_{b}\\
&&=I_{signature}(a)I_{flavor}(a)I_{spin+space}(a)+I_{signature}(b)I_{flavor}(b)I_{spin+space}(b)
\end{eqnarray}
Here $a$ and $b$ represent two decay diagrams in which the produced
quark goes into meson $B$ and meson $C$ respectively. In the present
work, only the first diagram is allowed. Therefore we have the
flavor factor $I_{flavor}(a)=1$ and $I_{flavor}(b)=0$, as is listed
in Table \ref{tab:flavor}. The spin-space part of $h_{fi}$ is
explicitly given by
\begin{eqnarray}
\nonumber I_{spin+space}(a)&&=\int d^3\mathbf{k}\;\Psi_{n_AL_AM_{L_A}}(\mathbf{k}-\mathbf{P_B})\Psi^*_{n_BL_BM_{L_B}}(\mathbf{k}-r\mathbf{P_B})\\
\label{overlap}&&\times\Psi^*_{n_CL_CM_{L_C}}(\mathbf{k}-r\mathbf{P_B})g\frac{m_3}{E_3}[\bar{u}_{\mathbf{k}s_{q_3}}v_{\mathbf{-k}s_{\bar{q}_4}}]
\end{eqnarray}
where $r=\frac{m_{q}}{m_{q}+m_{b}}$, and $m_{q}$ denotes created quark
mass $m_{u}$, $m_{d}$ or $m_{s}$. $\Psi_{n_AL_AM_{L_A}}$ is the
wavefunction of the initial meson $A$ in momentum space, and
$\Psi_{n_BL_BM_{L_B}}$ and $\Psi_{n_CL_CM_{L_C}}$ are the
wavefunctions of the final state mesons $B$ and $C$ respectively.
Taking into account the phase space, we get the differential decay
rate
\begin{equation}
\label{8}\frac{d\Gamma_{A\rightarrow BC}}{d\Omega}=2\pi\frac{P
E_{B}E_{C}}{M_{A}}|h_{fi}|^{2}
\end{equation}
where $P$ is the momentum of the final state mesons in the rest
frame of meson $A$
\begin{equation}
\label{9}P=\sqrt{[M^2_A-(M_B+M_C)^2][M^2_A-(M_B-M_C)^2]}\Big/(2M_A).
\end{equation}
To compare with the experiments, we transform the amplitude $h_{fi}$
into the partial wave amplitude ${\cal M}_{LS}$ by the recoupling
calculation \cite{Jacob:1959at}, then the decay width is
\begin{equation}
\label{add2}\Gamma(A\rightarrow B+C)=2\pi\frac{P
E_{B}E_{C}}{M_{A}}\sum_{LS}|{\cal M}_{LS}|^2.
\end{equation}
Since we neglect mass splitting within the same isospin multiplet,
to sum over all channels, one should multiply the mass shift due to
a specific hadronic loop by the flavor factor ${\cal F}$ which is
listed in Table \ref{tab:flavor}.

\begin{table}[t]
\begin{center}
\begin{tabular}{|c|c|c|c|c|}
\hline\hline Generic Decay&Example & $I_{flavor}(a)$
&$I_{flavor}(b)$ &~~${\cal F}~~$
\\\hline\hline
$X \rightarrow B\bar{B}$& $X \rightarrow B^{+}+{B}^{-}$   & $1$              & $0$             &   $2$ \\
$X \rightarrow B^{*}\bar{B}$& $X \rightarrow B^{*+}+{B}^{-}$   & $1$              & $0$        &   $4$ \\
$X \rightarrow B^{*}\bar{B}^{*}$& $X \rightarrow B^{*+}+{B}^{*-}$   & $1$              & $0$   &   $2$ \\
$X \rightarrow B_{s}\bar{B}_{s}$& $X \rightarrow B_{s}^{0}+\bar{B}_{s}^{0}$   & $1$         & $0$   &   $1$ \\
$X \rightarrow B_{s}^{*}\bar{B}_{s}$& $X \rightarrow B_{s}^{*0}+\bar{B}_{s}^{0}$   & $1$    & $0$   &   $2$ \\
$X \rightarrow B_{s}^{*}\bar{B}_{s}^{*}$& $X \rightarrow B_{s}^{*0}+\bar{B}_{s}^{*0}$   & $1$  & $0$  & $1$ \\
\hline\hline
\end{tabular}
\caption{\label{tab:flavor}Relevant flavor weight factors for
bottomonium decay, where $|X\rangle=|b\bar b\rangle$.}
\end{center}
\end{table}

We take all spatial wavefunctions to be simple harmonic oscillator
forms with $\beta_A$ being the oscillator parameter of the inital
meson $A$ and $\beta_B=\beta_C$ for the final state mesons $B$ and $C$. It
turn out that the transition amplitude ${\cal M}_{LS}$ is
proportional to an overall Gaussian factor, it can be expressed as
\begin{equation}
{\cal
M}_{LS}=\frac{\gamma}{\sqrt[4]{\pi}}e^{{-\frac{P^2(r-1)^2}{2\beta_{A}^2+\beta_{B}^2}}}{\cal
A}_{LS}
\end{equation}
where $r=\frac{m_{q}}{m_{q}+m_{b}}$, $m_{q}$ denotes created quark
mass $m_{u}$, $m_{d}$ or $m_{s}$, and $P$ is the momentum of the
final state mesons in the rest frame of meson $A$. The analytical
expressions of the amplitudes for the decays into two stable
$S-$wave final states are listed in the following.
We note that our expressions are different from the results in Ref.\cite{Barnes:1996ff}. The mass difference between the created quark and bottom quark and two oscillator parameters corresponding to initial and final states are considered in our expressions, and our results are more general than those in Ref.\cite{Barnes:1996ff}. In the limit of $\beta_A=\beta_B=\beta$ and $r=1/2$, the amplitudes presented below coincide with those of  Ref.\cite{Barnes:1996ff}.

\fbox{\bf 1S $\to$ 1S + 1S}
\begin{equation}
\mathrm{M_P} = \frac{-8P
\beta_{A}^{3/2}(2r\beta_{A}^2+\beta_{B}^2)}{\sqrt{3}(2\beta_{A}^2+\beta_{B}^2)^{5/2}}
\end{equation}
\begin{eqnarray}
  & & \nonumber \\ \fbox{$^3${\bf S}$_1$} & & \nonumber \\  & & \nonumber \\
 & {\cal A}_{10}{ ( ^3{\rm S}_1 \to ^1{\rm S}_0 + ^1{\rm S}_0 )} \  & =
\begin{array}{lr}
\mathrm{M_P}
\ &   ^1{\rm P}_1   \\
\end{array}
\\
& {\cal A}_{11}{ ( ^3{\rm S}_1 \to ^3{\rm S}_1 + ^1{\rm S}_0 )} \ &
=
\begin{array}{lr}
-\sqrt{2} \; \mathrm{M_P}
\ &   ^3{\rm P}_1   \\
\end{array}
\\
 & {\cal A}_{\rm LS}{ ( ^3{\rm S}_1 \to ^3{\rm S}_1 + ^3{\rm S}_1 )} \  & =
\left\{
\begin{array}{lr}
\sqrt{1\over 3 }\; \mathrm{M_P} \;
\ &   ^1{\rm P}_1   \\
0 \;
\ &   ^3{\rm P}_1   \\
-\sqrt{20 \over 3} \; \mathrm{M_P} \;
\ &   ^5{\rm P}_1   \\
0 \;
\ &   ^5{\rm F}_1   \\
\end{array}
\right.
\\
  & & \nonumber \\ \fbox{$^1${\bf S}$_0$} & & \nonumber \\  & & \nonumber \\
 & {\cal A}_{\rm LS}{ ( ^1{\rm S}_0 \to ^1{\rm S}_0 + ^1{\rm S}_0 )} \  & =
 0 \;
\\
 & {\cal A}_{11}{ ( ^1{\rm S}_0 \to ^3{\rm S}_1 + ^1{\rm S}_0 )} \  & =
\begin{array}{lr}
-\sqrt{3} \; \mathrm{M_P}
\ &   ^3{\rm P}_0   \\
\end{array}
\\
 & {\cal A}_{11}{ ( ^1{\rm S}_0 \to ^3{\rm S}_1 + ^3{\rm S}_1 )} \  & =
\begin{array}{lr}
\sqrt{6} \; \mathrm{M_P}
\ &   ^3{\rm P}_0   \\
\end{array}
\end{eqnarray}

\fbox{\bf 1P $\to$ 1S + 1S}

\begin{eqnarray}
 & \mathrm{M_S}  = -\frac{16
\beta_{A}^{3/2}(2(2r^2\beta_{A}^2-\beta_{B}^2+r(\beta_{B}^2-2\beta_{A}^2))P^2+3\beta_{B}^2(2\beta_{A}^2+\beta_{B}^2))}
{3\sqrt{\frac{2}{\beta_{B}^2}+\frac{1}{\beta_{A}^2}}\beta_{B}(2\beta_{A}^2+\beta_{B}^2)^{3}}
\\
 & \mathrm{M_D}  =
 \frac{32P^2(r-1)
\beta_{A}^{3/2}(2r\beta_{A}^2+\beta_{B}^2)}
{\sqrt{15}\sqrt{\frac{2}{\beta_{B}^2}+\frac{1}{\beta_{A}^2}}\beta_{B}(2\beta_{A}^2+\beta_{B}^2)^{3}}
\end{eqnarray}

\begin{eqnarray}
  & & \nonumber \\ \fbox{$^3${\bf P}$_2$} & & \nonumber \\  & & \nonumber \\
 & {\cal A}_{20}{ ( ^3{\rm P}_2 \to ^1{\rm S}_0 + ^1{\rm S}_0 )} \  & =
\mathrm{M_D} \;
\\
 & {\cal A}_{21}{( ^3{\rm P}_2 \to ^3{\rm S}_1 + ^1{\rm S}_0 ) }\  & =
-\sqrt{3\over 2} \; \mathrm{M_D} \;
\\
 & {\cal A}_{\rm LS}{( ^3{\rm P}_2 \to ^3{\rm S}_1 + ^3{\rm S}_1 ) }\  & =
\left\{
\begin{array}{cr}
-\sqrt{2}\; \mathrm{M_S} \;
\ &   ^5{\rm S}_2   \\
\sqrt{1\over 3} \; \mathrm{M_D} \;
\ &   ^1{\rm D}_2  \\
-\sqrt{7\over 3} \; \mathrm{M_D} \;
\ &   ^5{\rm D}_2  \\
\end{array}
\right. \\
  & & \nonumber \\ \fbox{$^3${\bf P}$_1$} & & \nonumber \\  & & \nonumber \\
 & {\cal A}_{\rm LS}{( ^3{\rm P}_1 \to ^3{\rm S}_1 + ^1{\rm S}_0 ) }\  & =
\left\{
\begin{array}{cr}
\mathrm{M_S}  \;
\ &   ^3{\rm S}_1   \\
-\sqrt{5\over 6} \; \mathrm{M_D} \;
\ &   ^3{\rm D}_1   \\
\end{array}
\right.
\\
 & {\cal A}_{\rm LS}{( ^3{\rm P}_1 \to ^3{\rm S}_1 + ^3{\rm S}_1 ) }\  & =
\left\{
\begin{array}{cr}
0 \;
\ &   ^3{\rm S}_1   \\
0 \;
\ &   ^3{\rm D}_1   \\
-\sqrt{5} \; \mathrm{M_D} \;
\ &   ^5{\rm D}_1   \\
\end{array}
\right. \\
  & & \nonumber \\ \fbox{$^3${\bf P}$_0$} & & \nonumber \\  & & \nonumber \\
 & {\cal A}_{00}{( ^3{\rm P}_0 \to ^1{\rm S}_0 + ^1{\rm S}_0 ) }\  & =
\begin{array}{cr}
\sqrt{3\over 2} \; \mathrm{M_S} \;
\ &   ^1{\rm S}_0   \\
\end{array}
\\
& {\cal A}_{\rm LS}{( ^3{\rm P}_0 \to ^3{\rm S}_1 + ^3{\rm S}_1 ) }\
& = \left\{
\begin{array}{cr}
\sqrt{1\over 2} \; \mathrm{M_S} \;
\ &   ^1{\rm S}_0   \\
-\sqrt{20\over 3} \; \mathrm{M_D} \;
\ &   ^5{\rm D}_0   \\
\end{array}
\right. \\
  & & \nonumber \\ \fbox{$^1${\bf P}$_1$} & & \nonumber \\  & & \nonumber \\
 & {\cal A}_{\rm LS}{( ^1{\rm P}_1 \to ^3{\rm S}_1 + ^1{\rm S}_0 ) }\  & =
\left\{
\begin{array}{cr}
-\sqrt{1\over 2} \; \mathrm{M_S} \;
\ &   ^3{\rm S}_1   \\
-\sqrt{5\over 3} \; \mathrm{M_D} \;
\ &   ^3{\rm D}_1   \\
\end{array}
\right.
\\
 & {\cal A}_{\rm LS}{( ^1{\rm P}_1 \to ^3{\rm S}_1 + ^3{\rm S}_1 ) }\  & =
\left\{
\begin{array}{cr}
\mathrm{M_S} \;
\ &   ^3{\rm S}_1   \\
\sqrt{10 \over 3} \; \mathrm{M_D} \;
\ &   ^3{\rm D}_1   \\
0 \;
\ &   ^5{\rm D}_1   \\
\end{array}
\right.
\end{eqnarray}

\fbox{\bf 1D $\to$ 1S + 1S}

\begin{eqnarray}
 &\mathrm{M_P}  &=-\frac{64\sqrt{\frac{2}{3}}P(r-1)
\beta_{A}^{5/2}(2(2r^2\beta_{A}^2-\beta_{B}^2+r(\beta_{B}^2-2\beta_{A}^2))P^2+5\beta_{B}^2(2\beta_{A}^2+\beta_{B}^2))}
{5\sqrt{\frac{2}{\beta_{B}^2}+\frac{1}{\beta_{A}^2}}\beta_{B}(2\beta_{A}^2+\beta_{B}^2)^4}
\\
 &\mathrm{M_F}
 &=-\frac{64P^3(r-1)^2
\beta_{A}^{5/2}(2r\beta_{A}^2+\beta_{B}^2)}
{\sqrt{35}\sqrt{\frac{2}{\beta_{B}^2}+\frac{1}{\beta_{A}^2}}\beta_{B}(2\beta_{A}^2+\beta_{B}^2)^4}
\end{eqnarray}

\begin{eqnarray}
  & & \nonumber \\ \fbox{$^3${\bf D}$_3$} & & \nonumber \\  & & \nonumber \\
 & {\cal A}_{30}{ ( ^3{\rm D}_3 \to ^1{\rm S}_0 + ^1{\rm S}_0 )} \  & =
\begin{array}{cr}
\mathrm{M_F} \;
\ &   ^1{\rm F}_3  \\
\end{array}
\\
 & {\cal A}_{31}{( ^3{\rm D}_3 \to ^3{\rm S}_1 + ^1{\rm S}_0 ) }\  & =
\begin{array}{cr}
-\sqrt{4\over 3} \; \mathrm{M_F} \;
\ &   ^3{\rm F}_3  \\
\end{array}
\\
 & {\cal A}_{\rm LS}{( ^3{\rm D}_3 \to ^3{\rm S}_1 + ^3{\rm S}_1 ) }\  & =
\left\{
\begin{array}{cr}
 \mathrm{M_P} \;
\ &   ^5{\rm P}_3   \\
\sqrt{1\over 3} \; \mathrm{M_F} \;
\ &   ^1{\rm F}_3  \\
0 \;
\ &   ^3{\rm F}_3  \\
-\sqrt{8\over 5} \; \mathrm{M_F} \;
\ &   ^5{\rm F}_3  \\
0 \;
\ &   ^5{\rm H}_3  \\
\end{array}
\right. \\
  & & \nonumber \\ \fbox{$^3${\bf D}$_2$} & & \nonumber \\  & & \nonumber \\
 & {\cal A}_{\rm LS}{( ^3{\rm D}_2 \to ^3{\rm S}_1 + ^1{\rm S}_0 ) }\  & =
\left\{
\begin{array}{cr}
-\sqrt{3\over 8} \; \mathrm{M_P}  \;
\ &   ^3{\rm P}_2   \\
-\sqrt{14\over 15} \; \mathrm{M_F} \;
\ &   ^3{\rm F}_2   \\
\end{array}
\right.  \\
 & {\cal A}_{\rm LS}{( ^3{\rm D}_2 \to ^3{\rm S}_1 + ^3{\rm S}_1 ) }\  & =
\left\{
\begin{array}{cr}
{1\over 2} \; \mathrm{M_P}  \;
\ &   ^5{\rm P}_2   \\
0\;
\ &   ^3{\rm F}_2   \\
-\sqrt{56\over 15} \; \mathrm{M_F} \;
\ &   ^5{\rm F}_2   \\
\end{array}
\right.  \\
  & & \nonumber \\ \fbox{$^3${\bf D}$_1$} & & \nonumber \\  & & \nonumber \\
 & {\cal A}_{10}{( ^3{\rm D}_1 \to ^1{\rm S}_0 + ^1{\rm S}_0 ) }\  & =
\begin{array}{cr}
-\sqrt{5\over 12} \; \mathrm{M_P} \;
\ &   ^1{\rm P}_1   \\
\end{array}
\\
 & {\cal A}_{11}{( ^3{\rm D}_1 \to ^3{\rm S}_1 + ^1{\rm S}_0 ) }\  & =
\begin{array}{cr}
-\sqrt{5\over 24} \; \mathrm{M_P} \;
\ &   ^3{\rm P}_1   \\
\end{array}
\\
& {\cal A}_{\rm LS}{( ^3{\rm D}_1 \to ^3{\rm S}_1 + ^3{\rm S}_1 ) }\
& = \left\{
\begin{array}{cr}
-{\sqrt{5}\over 6} \; \mathrm{M_P} \;
\ &   ^1{\rm P}_1   \\
0\;
\ &   ^3{\rm P}_1   \\
{1\over 6} \; \mathrm{M_P} \;
\ &   ^5{\rm P}_1   \\
-\sqrt{28\over 5} \; \mathrm{M_F} \;
\ &   ^5{\rm F}_1   \\
\end{array}
\right. \\
  & & \nonumber \\ \fbox{$^1${\bf D}$_2$} & & \nonumber \\  & & \nonumber \\
 & {\cal A}_{\rm LS}{( ^1{\rm D}_2 \to ^3{\rm S}_1 + ^1{\rm S}_0 ) }\  & =
\left\{
\begin{array}{cr}
{1\over 2} \; \mathrm{M_P} \;
\ &   ^3{\rm P}_2   \\
-\sqrt{7\over 5} \; \mathrm{M_F} \;
\ &   ^3{\rm F}_2   \\
\end{array}
\right.  \\
 & {\cal A}_{\rm LS}{( ^1{\rm D}_2 \to ^3{\rm S}_1 + ^3{\rm S}_1 ) }\  & =
\left\{
\begin{array}{cr}
-\sqrt{1\over 2} \; \mathrm{M_P} \;
\ &   ^3{\rm P}_2   \\
0\;
\ &   ^5{\rm P}_2   \\
\sqrt{14\over 5} \; \mathrm{M_F} \;
\ &   ^3{\rm F}_2   \\
0\;
\ &   ^5{\rm F}_2   \\
\end{array}
\right.
\end{eqnarray}

\fbox{\bf 1F $\to$ 1S + 1S}

\begin{eqnarray}
 &\mathrm{M_D}&=-\frac{128\sqrt{\frac{2}{5}}P^2(r-1)^2
\beta_{A}^{9/2}(2(2r^2\beta_{A}^2-\beta_{B}^2+r(\beta_{B}^2-2\beta_{A}^2))P^2+7\beta_{B}^2(2\beta_{A}^2+\beta_{B}^2))}
{7(2\beta_{A}^2+\beta_{B}^2)^{11/2}}
\\
 &\mathrm{M_G}&=-\frac{256\sqrt{\frac{2}{105}}P^4(r-1)^3
\beta_{A}^{9/2}(2r\beta_{A}^2+\beta_{B}^2)}
{3(2\beta_{A}^2+\beta_{B}^2)^{11/2}}
\end{eqnarray}

\begin{eqnarray}
  & & \nonumber \\ \fbox{$^3${\bf F}$_4$} & & \nonumber \\  & & \nonumber \\
 & {\cal A}_{40}{ ( ^3{\rm F}_4 \to ^1{\rm S}_0 + ^1{\rm S}_0 )} \  & =
\begin{array}{cr}
\mathrm{M_G} \;
\    &^1{\rm G}_4   \\
\end{array}
\\
 & {\cal A}_{41}{( ^3{\rm F}_4 \to ^3{\rm S}_1 + ^1{\rm S}_0 ) }\  & =
\begin{array}{cr}
-\sqrt{5\over 4} \; \mathrm{M_G} \;
\    &^3{\rm G}_4   \\
\end{array}
\\
 & {\cal A}_{\rm LS}{( ^3{\rm F}_4 \to ^3{\rm S}_1 + ^3{\rm S}_1 ) }\  & =
\left\{
\begin{array}{cr}
 \mathrm{M_D} \;
\ &   ^5{\rm D}_4   \\
\sqrt{1\over 3} \; \mathrm{M_G} \;
\ &   ^1{\rm G}_4  \\
0 \;
\ &   ^3{\rm G}_4   \\
-\sqrt{55\over 42} \; \mathrm{M_G} \;
\ &   ^5{\rm G}_4   \\
0 \;
\ &   ^5{\rm I}_4   \\
\end{array}
\right.  \\
  & & \nonumber \\ \fbox{$^3${\bf F}$_3$} & & \nonumber \\  & & \nonumber \\
 & {\cal A}_{\rm LS}{( ^3{\rm F}_3 \to ^3{\rm S}_1 + ^1{\rm S}_0 ) }\  & =
\left\{
\begin{array}{cr}
-\sqrt{1\over 3} \; \mathrm{M_D} \;
\ &   ^3{\rm D}_3   \\
-\sqrt{27\over 28} \; \mathrm{M_G} \;
\ &   ^3{\rm G}_3  \\
\end{array}
\right.
\\
 & {\cal A}_{\rm LS}{( ^3{\rm F}_3 \to ^3{\rm S}_1 + ^3{\rm S}_1 ) }\  & =
\left\{
\begin{array}{cr}
0 \;
\ &   ^3{\rm D}_3   \\
\sqrt{1\over 3} \;  \mathrm{M_D} \;
\ &   ^5{\rm D}_3   \\
0 \;
\ &   ^3{\rm G}_3   \\
-\sqrt{45\over 14} \; \mathrm{M_G} \;
\ &   ^5{\rm G}_3  \\
\end{array}
\right. \\
  & & \nonumber \\ \fbox{$^3${\bf F}$_2$} & & \nonumber \\  & & \nonumber \\
 & {\cal A}_{20}{ ( ^3{\rm F}_2 \to ^1{\rm S}_0 + ^1{\rm S}_0 )} \  & =
\begin{array}{cr}
- \sqrt{7\over 20} \; \mathrm{M_D} \;
\ &   ^1{\rm D}_2  \\
\end{array}
\\
 & {\cal A}_{21}{( ^3{\rm F}_2 \to ^3{\rm S}_1 + ^1{\rm S}_0 ) }\  & =
\begin{array}{cr}
-\sqrt{7\over 30} \; \mathrm{M_D} \;
\ &   ^3{\rm D}_2  \\
\end{array}
\\
 & {\cal A}_{\rm LS}{( ^3{\rm F}_2 \to ^3{\rm S}_1 + ^3{\rm S}_1 ) }\  & =
\left\{
\begin{array}{cr}
0 \;
\ &   ^5{\rm S}_2   \\
-\sqrt{7\over 60} \;  \mathrm{M_D} \;
\ &   ^1{\rm D}_2   \\
0 \;
\ &   ^3{\rm D}_2   \\
\sqrt{1\over 15} \;  \mathrm{M_D} \;
\ &   ^5{\rm D}_2   \\
-\sqrt{36\over 7} \; \mathrm{M_G} \;
\ &   ^5{\rm G}_2  \\
\end{array}
\right.  \\
  & & \nonumber \\ \fbox{$^1${\bf F}$_3$} & & \nonumber \\  & & \nonumber \\
 & {\cal A}_{\rm LS}{( ^1{\rm F}_3 \to ^3{\rm S}_1 + ^1{\rm S}_0 ) }\  & =
\left\{
\begin{array}{cr}
{1\over 2} \; \mathrm{M_D} \;
\ &   ^3{\rm D}_3   \\
-\sqrt{9\over 7} \; \mathrm{M_G} \;
\ &   ^3{\rm G}_3  \\
\end{array}
\right.
\\
 & {\cal A}_{\rm LS}{( ^1{\rm F}_3 \to ^3{\rm S}_1 + ^3{\rm S}_1 ) }\  & =
\left\{
\begin{array}{cr}
-\sqrt{1\over 2} \;  \mathrm{M_D} \;
\ &   ^3{\rm D}_3   \\
0 \;
\ &   ^5{\rm D}_3   \\
\sqrt{18\over 7} \; \mathrm{M_G} \;
\ &   ^3{\rm G}_3  \\
0 \;
\ &   ^5{\rm G}_3   \\
\end{array}
\right.
\end{eqnarray}

\fbox{\bf 1G $\to$ 1S + 1S}

\begin{eqnarray}
 &\mathrm{M_F}&=\frac{512P^3(r-1)^3
\beta_{A}^{11/2}(2(2r^2\beta_{A}^2-\beta_{B}^2+r(\beta_{B}^2-2\beta_{A}^2))P^2+9\beta_{B}^2(2\beta_{A}^2+\beta_{B}^2))}
{21\sqrt{15}\sqrt[4]{\pi}(2\beta_{A}^2+\beta_{B}^2)^{13/2}}
\\
 &\mathrm{M_H}&=-\frac{512P^5(r-1)^4
\beta_{A}^{11/2}(2r\beta_{A}^2+\beta_{B}^2)}
{3\sqrt{231}\sqrt[4]{\pi}(2\beta_{A}^2+\beta_{B}^2)^{13/2}}
\end{eqnarray}

\begin{eqnarray}
  & & \nonumber \\ \fbox{$^3${\bf G}$_5$} & & \nonumber \\  & & \nonumber \\
 & {\cal A}_{50}{ ( ^3{\rm G}_5 \to ^1{\rm S}_0 + ^1{\rm S}_0 )} \  & =
\begin{array}{cr}
\mathrm{M_H} \;
\    &^1{\rm H}_5   \\
\end{array}
\\
& {\cal A}_{51}{ ( ^3{\rm G}_5 \to ^1{\rm S}_0 + ^3{\rm S}_1 )} \ &
=
\begin{array}{cr}
\sqrt{6\over5} \;\mathrm{M_H} \;
\    &^3{\rm H}_5   \\
\end{array}
\\
 & {\cal A}_{\rm LS}{( ^3{\rm G}_5 \to ^3{\rm S}_1 + ^3{\rm S}_1 ) }\  & =
\left\{
\begin{array}{cr}
-\frac{2\sqrt{7}}{3} \;  \mathrm{M_F} \;
\ &   ^5{\rm F}_5   \\
\sqrt{1\over 3} \;  \mathrm{M_H} \;
\ &   ^1{\rm H}_5   \\
0 \;
\ &   ^3{\rm H}_5   \\
-\frac{2}{3}\sqrt{13\over 5} \; \mathrm{M_H} \;
\ &   ^5{\rm H}_5  \\
0 \;
\ &   ^5{\rm J}_5   \\
\end{array}
\right.
\end{eqnarray}

\begin{eqnarray}
  & & \nonumber \\ \fbox{$^3${\bf G}$_4$} & & \nonumber \\  & & \nonumber \\
 & {\cal A}_{LS}{ ( ^3{\rm G}_4 \to ^1{\rm S}_0 + ^3{\rm S}_1 )} \ &=
\left\{
\begin{array}{cr}
\frac{-\sqrt{35}}{6} \;\mathrm{M_F} \;
\    &^3{\rm F}_4   \\
\frac{2\sqrt{11}}{3\sqrt{5}} \;\mathrm{M_H} \;
\    &^3{\rm H}_4   \\
\end{array}
\right.
\\
 & {\cal A}_{\rm LS}{( ^3{\rm G}_4 \to ^3{\rm S}_1 + ^3{\rm S}_1 ) }\  & =
\left\{
\begin{array}{cr}
0 \;
\ &   ^3{\rm F}_4   \\
0 \;
\ &   ^3{\rm H}_4   \\
-\sqrt{\frac{7}{6}} \; \mathrm{M_F} \;
\ &   ^5{\rm F}_4   \\
-2\sqrt{\frac{11}{15}} \; \mathrm{M_H} \;
\ &   ^5{\rm H}_4  \\
\end{array}
\right.
\end{eqnarray}

\begin{eqnarray}
  & & \nonumber \\ \fbox{$^3${\bf G}$_3$} & & \nonumber \\  & & \nonumber \\
   & {\cal A}_{30}{ ( ^3{\rm G}_3 \to ^1{\rm S}_0 + ^1{\rm S}_0 )} \ &
=
\begin{array}{cr}
\mathrm{M_F} \;
\    &^1{\rm F}_3   \\
\end{array}
\\
 & {\cal A}_{31}{ ( ^3{\rm G}_3 \to ^1{\rm S}_0 + ^3{\rm S}_1 )} \ &
=
\begin{array}{cr}
\frac{-\sqrt{3}}{2} \;\mathrm{M_F} \;
\    &^3{\rm F}_3   \\
\end{array}
\\
 & {\cal A}_{\rm LS}{( ^3{\rm G}_3 \to ^3{\rm S}_1 + ^3{\rm S}_1 ) }\  & =
\left\{
\begin{array}{cr}
\frac{1}{\sqrt{3}} \; \mathrm{M_F} \;
\ &   ^1{\rm F}_3   \\
0 \;
\ &   ^3{\rm F}_3   \\
0 \;
\ &   ^5{\rm P}_3   \\
-\frac{\sqrt{5}}{3\sqrt{2}}\;\mathrm{M_F} \;
\ &   ^5{\rm F}_3   \\
-\frac{2\sqrt{11}}{3}\; \mathrm{M_H} \;
\ &   ^5{\rm H}_3  \\
\end{array}
\right.
\end{eqnarray}

\begin{eqnarray}
  & & \nonumber \\ \fbox{$^1${\bf G}$_4$} & & \nonumber \\  & & \nonumber \\
   & {\cal A}_{LS}{ ( ^1{\rm G}_4 \to ^1{\rm S}_0 + ^3{\rm S}_1 )} \ &
=\left\{
\begin{array}{cr}
\frac{-\sqrt{7}}{3} \;\mathrm{M_F} \;
\    &^3{\rm F}_4   \\
\frac{\sqrt{11}}{3} \;\mathrm{M_H} \;
\    &^3{\rm H}_4   \\
\end{array}
\right.
\\
 & {\cal A}_{\rm LS}{( ^1{\rm G}_4 \to ^3{\rm S}_1 + ^3{\rm S}_1 ) }\  & =
\left\{
\begin{array}{cr}
\frac{\sqrt{14}}{3} \; \mathrm{M_F} \;
\ &   ^3{\rm F}_4   \\
\frac{\sqrt{22}}{3} \; \mathrm{{M_H}}  \;
\ &   ^3{\rm H}_4   \\
0 \;
\ &   ^5{\rm F}_4   \\
0 \;
\ &   ^5{\rm H}_4 \\
\end{array}
\right.
\end{eqnarray}

\begin{center}
{\bf Appendix B: Recursion relations between wavefunctions}
\end{center}

In Appendix A, we present the amplitudes for the ground bottomonium
states decaying into two $S-$wave final states. If the initial state
is the radial excited bottomonium, the corresponding amplitudes can
certainly be derived in the same way. However, it is interesting to
notice that the radially excited wavefunctions can be related to
the lowest radial wavefunctions by differentiation,
\begin{equation}
\Psi_{1S}(\mathbf{p})=\frac{1}{\beta^{3/2}\pi^{3/4}}e^{-\frac{p^2}{2\beta^2}}
\end{equation}

\begin{equation}
\Psi_{2S}(\mathbf{p})=\frac{1}{\sqrt{6}\beta^2}(-3\beta^2+2p^{2})\Psi_{1S}(\mathbf{p})=\frac{2}{\sqrt{6}}\beta\frac{\partial}{\partial
\beta}\Psi_{1S}(\mathbf{p})
\end{equation}

\begin{equation}
\Psi_{3S}(\mathbf{p})=\frac{1}{2}\sqrt{\frac{15}{2}}(\frac{4p^{4}}{15\beta^4}-\frac{4p^{2}}{3\beta^2}+1)\Psi_{1S}(\mathbf{p})=\frac{1}{\sqrt{30}}(3+2\beta\frac{\partial}{\partial
\beta}+2\beta^2\frac{\partial^2}{\partial\beta^2})\Psi_{1S}(\mathbf{p})
\end{equation}

\begin{eqnarray}
\Psi_{4S}(\mathbf{p})&&=\frac{-1}{4}\sqrt{35}(\frac{-8p^{6}}{105\beta^6}+\frac{4p^{4}}{5\beta^4}-2\frac{p^{2}}{\beta^2}+1)\Psi_{1S}(\mathbf{p})
\nonumber\\
&&=\frac{1}{3 \sqrt{35}}(15\beta\frac{\partial}{\partial
\beta}+6\beta^2\frac{\partial^2}{\partial\beta^2}+2\beta^3\frac{\partial^3}{\partial\beta^3})\Psi_{1S}(\mathbf{p})
\end{eqnarray}

\begin{eqnarray}
\Psi_{5S}(\mathbf{p})&&=\frac{3}{8}\sqrt{\frac{35}{2}}(\frac{16p^{8}}{945\beta^8}+\frac{-32p^{6}}{105\beta^6}+\frac{8p^{4}}{5\beta^4}+\frac{-8p^{2}}{3\beta^2}+1)\Psi_{1S}(\mathbf{p})
\nonumber\\
&&=\frac{1}{18 \sqrt{70}}(63+72\beta\frac{\partial}{\partial
\beta}+96\beta^2\frac{\partial^2}{\partial\beta^2}+24\beta^3\frac{\partial^3}{\partial\beta^3}+4\beta^4\frac{\partial^4}{\partial
\beta^4})\Psi_{1S}(\mathbf{p})
\end{eqnarray}

\begin{eqnarray}
&&\Psi_{6S}(\mathbf{p})=\frac{-3\sqrt{77}}{16}(-\frac{32p^{10}}{10395\beta^{10}}+\frac{16p^{8}}{189\beta^8}+\frac{-16p^{6}}{21\beta^6}+\frac{8p^{4}}{3\beta^4}+\frac{-10p^{2}}{3\beta^2}+1)\Psi_{1S}(\mathbf{p})
\nonumber\\
&&=\frac{1}{45 \sqrt{77}}(\frac{675}{2}\beta\frac{\partial}{\partial
\beta}+240\beta^2\frac{\partial^2}{\partial\beta^2}+120\beta^3\frac{\partial^3}{\partial\beta^3}
+20\beta^4\frac{\partial^4}{\partial\beta^4}+2\beta^5\frac{\partial^5}{\partial\beta^5})\Psi_{1S}(\mathbf{p})
\end{eqnarray}

\begin{equation}
\Psi_{2P}(\mathbf{p})=4\sqrt{\frac{2!}{5!}}\frac{1}{\pi^{1/4}\beta^{5}}p(-5+\frac{2p^2}{\beta^2})Y_{1M_{L}}(\widehat{\mathbf{p}})e^{-\frac{p^2}{2\beta^2}}
=\sqrt{\frac{2}{5}}\beta\frac{\partial}{\partial
\beta}\Psi_{1P}(\mathbf{p})
\end{equation}

\begin{eqnarray}
\Psi_{3P}(\mathbf{p})=\frac{1}{2\sqrt{70}}(\frac{4p^4}{\beta^4}-\frac{28p^2}{\beta^2}+35)\Psi_{1P}(\mathbf{p})
=\frac{1}{\sqrt{70}}(5+2\beta\frac{\partial}{\partial
\beta}+2\beta^2\frac{\partial^2}{\partial\beta^2})\Psi_{1P}(\mathbf{p})
\end{eqnarray}

\begin{equation}
\Psi_{2D}(\mathbf{p})=8\sqrt{\frac{3!}{7!}}\frac{1}{\pi^{1/4}\beta^{7}}p^{2}(-7+\frac{2p^2}{\beta^2})Y_{2M_{L}}(\widehat{\mathbf{p}})e^{-\frac{p^2}{2\beta^2}}
=\sqrt{\frac{2}{7}}\beta\frac{\partial}{\partial
\beta}\Psi_{1D}(\mathbf{p})
\end{equation}

\begin{eqnarray}
\Psi_{3D}(\mathbf{p})=\frac{1}{6\sqrt{14}}(\frac{4p^4}{\beta^4}-\frac{36p^2}{\beta^2}+63)\Psi_{1D}(\mathbf{p})
=\frac{1}{3\sqrt{14}}(7+2\beta\frac{\partial}{\partial
\beta}+2\beta^2\frac{\partial^2}{\partial\beta^2})\Psi_{1D}(\mathbf{p})
\end{eqnarray}

\begin{equation}
\Psi_{2F}(\mathbf{p})=\frac{1}{3\sqrt{2}}(-9+\frac{2p^2}{\beta^2})\Psi_{1F}(\mathbf{p})
=\frac{\sqrt{2}}{3}\beta\frac{\partial}{\partial
\beta}\Psi_{1F}(\mathbf{p})
\end{equation}
The differential operators depend only on $\beta$, hence they can be
pulled out of the integration in Eq.(\ref{overlap}). Therefore the
amplitudes for radially excited meson decays can be found by
applying the above differential operators to the amplitudes listed
in Appendix A. For example,
\begin{equation}
{\cal M}_{LS}(2\,^3{\rm S}_1\to ^1{\rm S}_0+ ^1{\rm
S}_0)=\frac{2}{\sqrt{6}}\beta_{A}\frac{\partial}{\partial
\beta_{A}}{\cal M}_{LS}(^3{\rm S}_1\to ^1{\rm S}_0+ ^1{\rm S}_0)
\end{equation}
We have checked that the amplitudes obtained with this method are
exactly the same as those that result from performing the overlap integral
straightforwardly.

\newpage

\begin{table}[hptb]
\begin{tabular}{|c|c|c|c||c|c|c|c|}
\hline\hline
states&$M_{ex}$&$M_{th}$&$M_{np}$&states&$M_{ex}$&$M_{th}$&$M_{np}$\\
\hline
$\eta_{b}(1\,^1S_0)$&9390.9&9391.8&9396.3&$\Upsilon(2\,{^3}{D}{_2})$&---&10426.8&10442.1\\
$\Upsilon(1\,{^3}{S}{_1})$&9460.3&9460.3&9444.2&$\Upsilon(2\,{^3}{D}{_3})$&---&10431.4&10447.0\\
$h_{b}(1\,{^1}{P}{_1})$&9899.9&9915.5&9907.9&$h_{b}(3\,{^1}{P}{_1})$&---&10523.2&10537.2\\
$\chi_{b0}(1\,{^3}{P}{_0})$&9859.4&9875.3&9869.1&$\chi_{b0}(3\,{^3}{P}{_0})$&---&10495.9&10508.1\\
$\chi_{b1}(1\,{^3}{P}{_1})$&9892.8&9906.8&9899.8&$\chi_{b1}(3\,{^3}{P}{_1})$&---&10517.3&10531.1\\
$\chi_{b2}(1\,{^3}{P}{_2})$&9912.2&9929.6&9921.7&$\chi_{b2}(3\,{^3}{P}{_2})$&---&10532.4&10547.9\\
$\eta_{b}(2\,{^1}{S}{_0})$&---&10004.9&9994.1&$\Upsilon(2\,{^1}{F}{_3})$&---&10566.4&10600.5\\
$\Upsilon(2\,^3S_1)$&10023.3&10026.2&10010.2&$\Upsilon(2\,{^3}{F}{_2})$&---&10560.9&10598.3\\
$\Upsilon(1\,{^1}{D}{_2})$&---&10145.5&10154.1&$\Upsilon(2\,{^3}{F}{_3})$&---&10566.1&10600.5\\
$\Upsilon(1\,{^3}{D}{_1})$&---&10138.1&10146.6&$\Upsilon(2\,{^3}{F}{_4})$&---&10567.9&10601.6\\
$\Upsilon(1\,{^3}{D}{_2})$&10164.5&10144.6&10153.2&$\eta_{b}(4\,{^1}{S}{_0})$&---&10593.2&10612.3\\
$\Upsilon(1\,{^3}{D}{_3})$&---&10149.3&10158.0&$\Upsilon(4\,{^3}{S}{_1})$&10579.4&10602.7&10621.0\\

$h_{b}(2\,{^1}{P}{_1})$&10259.8&10259.1&10257.7&$\Upsilon(3\,{^1}{D}{_2})$&---&10655.7&10693.0\\
$\chi_{b0}(2\,{^3}{P}{_0})$&10232.5&10227.9&10225.6&$\Upsilon(3\,{^3}{D}{_1})$&---&10650.9&10685.7\\
$\chi_{b1}(2\,{^3}{P}{_1})$&10255.5&10252.4&10251.0&$\Upsilon(3\,{^3}{D}{_2})$&---&10654.5&10692.0\\
$\chi_{b2}(2\,{^3}{P}{_2})$&10268.7&10270.1&10269.3&$\Upsilon(3\,{^3}{D}{_3})$&---&10662.8&10697.0\\

$\Upsilon(1\,{^1}{F}{_3})$&---&10322.1&10343.1&$\eta_{b}(5\,{^1}{S}{_0})$&---&10812.6&10852.3\\
$\Upsilon(1\,{^3}{F}{_2})$&---&10319.5&10341.3&$\Upsilon(5\,{^3}{S}{_1})$&10865.0&10819.9&10859.6\\
$\Upsilon(1\,{^3}{F}{_3})$&---&10322.0&10343.2&$\eta_{b}(6\,{^1}{S}{_0})$&---&11008.0&11070.0\\
$\Upsilon(1\,{^3}{F}{_4})$&---&10323.1&10344.0&$\Upsilon(6\,{^3}{S}{_1})$&11019.0&11022.6&11076.4\\

$\eta_{b}(3\,{^1}{S}{_0})$&---&10337.9&10337.5&$\Upsilon(1\,{^1}{G}{_4})$&---&10473.3&10505.4\\
$\Upsilon(3\,{^3}{S}{_1})$&10355.2&10351.9&10348.4&$\Upsilon(1\,{^3}{G}{_3})$&---&10471.8&10505.8\\
$\Upsilon(2\,{^1}{D}{_2})$&---&10427.98&10443.0&$\Upsilon(1\,{^3}{G}{_4})$&---&10473.4&10505.8\\
$\Upsilon(2\,{^3}{D}{_1})$&---&10420.4&10435.7&$\Upsilon(1\,{^3}{G}{_5})$&---&10470.8&10504.7\\

\hline\hline
\end{tabular}
\caption{\label{tab:allpredicted} The spectrum of the bottomonium
states, where $M_{ex}$ is the PDG average for the measured mass
\cite{pdg}. $M_{np}$ denotes the prediction for the mass of the
$b\bar{b}$ state in the conventional non-relativistic potential model
presented in section III. $M_{th}$ is the theoretical prediction
after the coupled-channel effects are taken into account. All the
masses are in units of megaelectronvolts (MeV).}
\end{table}

\begin{table}[hptb]
\begin{ruledtabular}
\small
\renewcommand{\arraystretch}{-1}
\addtolength{\tabcolsep}{-0.5pt}
\begin{tabular}{|c|cccccccccc|}
$\quad\quad$&$\eta_{b}(1S)$&$\Upsilon(1S)$&$h_{b}(1P)$&$\chi_{b0}(1P)$&$\chi_{b1}(1P)$&$\chi_{b2}(1P)$&$\Upsilon(2S)$&
$\Upsilon(1\,{^3}{D}_{2})$&$h_{b}(2P)$&$\chi_{b0}(2P)$\\
\hline
$BB$&0&4.209&0&15.523&0&8.588&5.164&0&0&15.305\\
$B^{*}B$&22.9806&16.228&36.892&0&33.483&24.364&19.418&40.776&34.508&0\\
$B^{*}B^{*}$&21.944&27.411&34.949&52.980&37.548&40.246&32.071&42.213&31.924&47.899\\
$B_{s}B_{s}$&0&0.908&0&2.528&0&1.727&1.017&0&0&2.332\\
$B_{s}^{*}B_{s}$&5.293&3.519&7.102&0&6.011&4.952&3.879&6.763&6.346&0\\
$B_{s}^{*}B_{s}^{*}$&5.281&5.971&6.797&10.795&7.738&7.456&6.488&7.941&6.003&9.510\\

\hline
$\delta M$&55.498&58.246&85.741&81.826&84.778&87.333&68.037&97.692&78.781&75.047\\
$M_{0}$&9447.3&9518.6&10001.2&9957.1&9991.6&10016.9&10094.2&10242.3&10337.9&10302.9\\
$M_{th}$&9391.8&9460.3&9915.5&9875.3&9906.8&9929.6&10026.2&10144.6&10259.1&10227.9\\
$M_{ex}$&9390.9&9460.3&9899.87&9859.44&9892.78&9912.21&10023.26&10164.5&10259.8&10232.5\\
$P_{b\bar{b}}$&0.877&0.959&0.910&0.917&0.912&0.907&0.923&0.870&0.885&0.894 \\
\hline\hline
$\quad\quad$&$\chi_{b1}(2P)$&$\chi_{b2}(2P)$&$\Upsilon(3S)$&$\Upsilon(4S)$&$\Upsilon(5S)$&$\Upsilon(6S)$&
$\eta_{b}(2S)$&$\eta_{b}(3S)$&$\eta_{b}(4S)$&$\eta_{b}(5S)$\\
\hline
$BB$&0&8.165&5.441&6.548&8.963&2.884&0&0&0&0\\
$B^{*}B$&31.697&22.568&19.828&24.377&28.159&21.871&28.813&28.9541&33.6972&39.649\\
$B^{*}B^{*}$&33.951&37.106&31.938&34.353&35.684&48.760&26.783&26.720&28.662&29.955\\
$B_{s}B_{s}$&0&1.554&0.992&1.014&1.048&1.145&0&0&0&0\\
$B_{s}^{*}B_{s}$&5.410&4.400&3.756&3.770&3.952&4.064&5.694&5.533&5.573&5.809\\
$B_{s}^{*}B_{s}^{*}$&6.804&6.611&6.215&6.178&6.424&6.729&5.447&5.238&5.224&5.410\\

\hline
$\delta M$&77.863&80.403&68.174&76.296&84.229&85.455&66.215&66.446&73.155&80.822\\
$M_{0}$&10330.3&10350.5&10420.1&10679.0&10904.2&11108.1&10071.26&10404.4&10666.4&10893.4\\
$M_{th}$&10252.4&10270.1&10351.9&10602.7&10819.9&11022.6&10004.9&10337.9&10593.2&10812.6\\
$M_{ex}$&10255.46&10268.65&10355.2&10579.4&10865.0&11019.0&---&---&---&---\\
$P_{b\bar{b}}$&0.887&0.881&0.893&0.638&0.791&0.891&0.926&0.898&0.776&0.780\\
\hline\hline
$\quad\quad$&$\eta_{b}(6S)$&$h_{b}(3P)$&$\chi_{b0}(3P)$&$\chi_{b1}(3P)$&$\chi_{b2}(3P)$&$\Upsilon(1\,{^1}{D}{_2})$&
$\Upsilon(1\,{^3}{D}{_1})$&$\Upsilon(1\,{^3}{D}{_3})$&$\Upsilon(2\,{^1}{D}{_2})$&$\Upsilon(2\,{^3}{D}{_1})$\\
\hline
$BB$&0&0&17.469&0&9.065&0&13.077&10.995&0&13.187\\
$B^{*}B$&36.924&36.078&0&33.683&23.284&43.057&11.968&27.296&40.150&11.383\\
$B^{*}B^{*}$&41.997&31.981&46.976&33.449&37.752&40.082&57.205&45.513&36.199&51.163\\
$B_{s}B_{s}$&0&0&2.291&0&1.499&0&1.817&2.088&0&1.573\\
$B_{s}^{*}B_{s}$&5.980&6.077&0&5.207&4.195&7.575&1.703&5.272&6.657&1.445\\
$B_{s}^{*}B_{s}^{*}$&5.611&5.688&8.980&6.423&6.282&7.175&11.023&7.531&6.234&9.663\\

\hline
$\delta M$&90.511&79.824&75.716&78.762&82.077&97.890&96.793&98.696&89.239&88.412\\
$M_{0}$&11098.5&10603.0&10571.6&10596.1&10614.5&10243.4&10234.9&10248.0&10517.1&10508.8\\
$M_{th}$&11008.0&10523.2&10495.9&10517.3&10532.4&10145.5&10138.1&10149.3&10427.9&10420.4\\
$M_{ex}$&---&---&---&---&---&---&---&---&---&---\\
$P_{b\bar{b}}$&0.883&0.831&0.843&0.835&0.816&0.870&0.872&0.869&0.836&0.835\\
\end{tabular}
\addtolength{\tabcolsep}{-0.5pt}
\renewcommand{\arraystretch}{-1}
\small
\end{ruledtabular}
\caption{\label{tab:shift1} Mass shifts (in MeV) and
$b\bar{b}$ component $P_{b\bar{b}}$ of bottomonium states due to
individual $B\bar{B}$ loops. $M_{ex}$ is the PDG average for the
measured mass, $M_0$ denotes the bare mass, and $M_{th}$ denotes the
predicted masses with coupled-channel effects considered. $\delta M$
denotes the total mass shift. For simplicity, we have abbreviated
the $B\bar{B}$ hadronic loop as $``BB"$, $B\bar{B}^{*}+B^{*}\bar{B}$
as $``BB^{*}"$, and so forth.}
\end{table}

\begin{table}[hptb]
\begin{ruledtabular}
\small
\renewcommand{\arraystretch}{-1}
\addtolength{\tabcolsep}{-0.5pt}
\begin{tabular}{|c|ccccccccc|}
$\quad\quad$&$\Upsilon(2\,{^3}{
D}{_2})$&$\Upsilon(2\,{^3}{D}{_3})$&$\Upsilon(3\,{^1}{D}{_2})$&$\Upsilon(3\,{^3}{D}{_1})$&$\Upsilon(3\,{^3}{D}{_2})$&
$\Upsilon(3\,{^3}{D}{_3})$&$\Upsilon(1\,{^1}{F}{_3})$&$\Upsilon(1\,{^3}{F}{_2})$&$\Upsilon(1\,{^3}{F}{_3})$\\
\hline
$BB$&0&10.460&0&13.300&0&11.295&0&13.733&0\\
$B^{*}B$&38.298&25.121&43.392&13.795&43.215&24.146&46.402&16.135&44.945\\
$B^{*}B^{*}$&37.940&41.427&41.000&54.099&41.229&45.943&42.246&58.809&43.726\\
$B_{s}B_{s}$&0&1.879&0&1.564&0&1.841&0&1.631&0\\
$B_{s}^{*}B_{s}$&5.892&4.690&6.396&1.397&5.664&4.531&7.435&2.001&6.831\\
$B_{s}^{*}B_{s}^{*}$&6.948&6.465&5.921&9.211&6.601&6.181&6.965&10.636&7.542\\

\hline
$\delta M$&89.078&90.041&96.710&93.367&96.709&93.936&103.048&102.945&103.043\\
$M_{0}$&10515.9&10521.5&10752.4&10744.3&10751.2&10756.7&10425.1&10422.5&10425.0\\
$M_{th}$&10426.8&10431.4&10655.7&10650.9&10654.5&10662.8&10322.1&10319.5&10322.0\\
$M_{ex}$&---&---&---&---&---&---&---&---&---\\
$P_{b\bar{b}}$&0.836&0.835&0.922&0.645&0.763&0.588&0.829&0.827&0.829\\
\hline\hline
$\quad\quad$&$\Upsilon(1\,{^3}{F}{_4})$&$\Upsilon(2\,{^1}{F}{_3})$&$\Upsilon(2\,{^3}{F}{_2})$&$\Upsilon(2\,{^3}{F}{_3})$&$\Upsilon(2\,{^3}{F}{_4})$&$\Upsilon(1\,{^1}{G}{_4})$&$\Upsilon(1\,{^3}{G}{_3})$&$\Upsilon(1\,{^3}{G}{_4})$&$\Upsilon(1\,{^3}{G}{_5})$\\
\hline
$BB$&12.378&0&18.347&0&13.133&0&15.860&0&15.784\\
$B^{*}B$&28.246&47.250&17.108&46.383&27.719&49.311&19.559&48.645&31.174\\
$B^{*}B^{*}$&48.263&40.546&55.203&41.679&47.068&43.356&58.613&44.229&50.179\\
$B_{s}B_{s}$&2.194&0&1.423&0&2.066&0&1.520&0&1.517\\
$B_{s}^{*}B_{s}$&5.142&6.757&1.699&6.140&4.789&7.005&2.050&6.565&3.272\\
$B_{s}^{*}B_{s}^{*}$&7.175&6.257&9.705&6.844&6.288&6.475&9.815&6.900&6.608\\

\hline
$\delta M$&103.398&100.810&103.484&101.047&101.064&106.147&107.417&106.338&108.533\\
$M_{0}$&10426.5&10667.2&10664.4&10667.1&10668.9&10579.4&10579.2&10579.7&10579.3\\
$M_{th}$&10323.1&10566.4&10560.9&10566.1&10567.9&10473.3&10471.8&10473.4&10470.8\\
$M_{ex}$&---&---&---&---&---&---&---&---&---\\
$P_{b\bar{b}}$&0.830&0.751&0.637&0.748&0.749&0.775&0.762&0.773&0.733\\
\end{tabular}
\addtolength{\tabcolsep}{-0.5pt}
\renewcommand{\arraystretch}{-1}
\small
\end{ruledtabular}
\caption{\label{tab:shift2}The continuing of Table
\ref{tab:shift1}.}
\end{table}


\begin{thebibliography}{99}

\bibitem{Swanson:2006st}
  E.~S.~Swanson,
  Phys.\ Rept.\  {\bf 429}, 243 (2006)
  [arXiv:hep-ph/0601110].

\bibitem{Eichten:2007qx}
  E.~Eichten, S.~Godfrey, H.~Mahlke and J.~L.~Rosner,
  Rev.\ Mod.\ Phys.\  {\bf 80}, 1161 (2008)
  [arXiv:hep-ph/0701208].

\bibitem{Godfrey:2008nc}
  S.~Godfrey and S.~L.~Olsen,
  Ann.\ Rev.\ Nucl.\ Part.\ Sci.\  {\bf 58}, 51 (2008)
  [arXiv:0801.3867 [hep-ph]].


\bibitem{Brambilla:2010cs}
  N.~Brambilla, S.~Eidelman, B.~K.~Heltsley, R.~Vogt, G.~T.~Bodwin, E.~Eichten, A.~D.~Frawley and A.~B.~Meyer {\it et al.},
  Eur.\ Phys.\ J.\ C {\bf 71}, 1534 (2011)
  [arXiv:1010.5827 [hep-ph]].


\bibitem{Aubert:2008a}
  B. Aubert {\it et al.} BABAR Collaboration, Phys.\ Rev.\ Lett.\
  {\bf 101}, 071801 (2008), [arXiv:0807.1086].

\bibitem{Sanchez:2010a}
  P. Sanchez {\it et al.} BABAR Collaboration, Phys.\ Rev.\ D.\
  {\bf 82}, 111102 (2010), [arXiv:1004.0175].

\bibitem{Lees:2011a}
  J. Lees {\it et al.} BABAR Collaboration, [arXiv:1102.4565].

\bibitem{Adachi:2011a}
  I. Adachi {\it et al.} Belle Collaboration, [arXiv:1103.3419].


\bibitem{Aad:2011ih}
  G.~Aad {\it et al.}  [ATLAS Collaboration],
  arXiv:1112.5154 [hep-ex].


\bibitem{:2008pu}
K. F. Chen et al. [Belle Collaboration], Phys. Rev. Lett.
100, 112001 (2008); I. Adachi et al. [Belle Collaboration], Phys. Rev. D 82,
091106 (2010), arXiv:0808.2445 [hep-ex].


\bibitem{:2008hx}
  B.~Aubert {\it et al.}  [BABAR Collaboration],
  Phys.\ Rev.\ Lett.\  {\bf 102}, 012001 (2009)
  [arXiv:0809.4120 [hep-ex]].

\bibitem{Collaboration:2011gja}
I. Adachi et.al. [Belle Collaboration] arXiv:1105.4583 [hep-ex].


\bibitem{LHCb}
LHCb homepage, http://lhcb.web.cern.ch/lhcb


\bibitem{O'Leary:2010af}
  B.~O'Leary {\it et al.}  [SuperB Collaboration],
  arXiv:1008.1541 [hep-ex].


\bibitem{Tornqvist:1979hx}
  N.~A.~Tornqvist,
  Annals Phys.\  {\bf 123}, 1 (1979);  N.~A.~Tornqvist,
  Acta Phys.\ Polon.\  B {\bf 16}, 503 (1985)
  [Erratum-ibid.\  B {\bf 16}, 683 (1985)].

\bibitem{Ono:1983rd}
  S.~Ono and N.~A.~Tornqvist,
  Z.\ Phys.\  C {\bf 23}, 59 (1984); K.~Heikkila, S.~Ono and N.~A.~Tornqvist,
  Phys.\ Rev.\  D {\bf 29}, 110 (1984)
  [Erratum-ibid.\  D {\bf 29}, 2136 (1984)].


\bibitem{Tornqvist:1995kr}
  N.~A.~Tornqvist,
  Z.\ Phys.\  C {\bf 68}, 647 (1995)
  [arXiv:hep-ph/9504372]; N.~A.~Tornqvist and M.~Roos,
  Phys.\ Rev.\ Lett.\  {\bf 76}, 1575 (1996)
  [arXiv:hep-ph/9511210].


\bibitem{vanBeveren:1979bd}
  E.~van Beveren, C.~Dullemond and G.~Rupp,
  Phys.\ Rev.\  D {\bf 21}, 772 (1980)
  [Erratum-ibid.\  D {\bf 22}, 787 (1980)]; E.~van Beveren, G.~Rupp, T.~A.~Rijken and C.~Dullemond,
  Phys.\ Rev.\  D {\bf 27}, 1527 (1983).


\bibitem{Eichten:1974af}
  E.~Eichten, K.~Gottfried, T.~Kinoshita, J.~B.~Kogut, K.~D.~Lane and T.~M.~Yan,
  Phys.\ Rev.\ Lett.\  {\bf 34}, 369 (1975)
  [Erratum-ibid.\  {\bf 36}, 1276 (1976)]; E.~Eichten, K.~Gottfried, T.~Kinoshita, K.~D.~Lane and T.~M.~Yan,
  Phys.\ Rev.\ Lett.\  {\bf 36}, 500 (1976); E.~Eichten, K.~Gottfried, T.~Kinoshita, K.~D.~Lane and T.~M.~Yan,
  Phys.\ Rev.\  D {\bf 17}, 3090 (1978)
  [Erratum-ibid.\  D {\bf 21}, 313 (1980)]; E.~Eichten, K.~Gottfried, T.~Kinoshita, K.~D.~Lane and T.~M.~Yan,
  Phys.\ Rev.\  D {\bf 21}, 203 (1980); E.~J.~Eichten, K.~Lane, C.~Quigg,
  Phys.\ Rev.\  {\bf D73}, 014014 (2006).
  [hep-ph/0511179].


\bibitem{Barnes:2003dj}
  T.~Barnes, F.~E.~Close and H.~J.~Lipkin,
  Phys.\ Rev.\  D {\bf 68}, 054006 (2003)
  [arXiv:hep-ph/0305025].


\bibitem{vanBeveren:2003kd}
  E.~van Beveren and G.~Rupp,
  Phys.\ Rev.\ Lett.\  {\bf 91}, 012003 (2003)
  [arXiv:hep-ph/0305035].

\bibitem{Hwang:2004cd}
  D.~S.~Hwang and D.~W.~Kim,
  Phys.\ Lett.\  B {\bf 601}, 137 (2004)
  [arXiv:hep-ph/0408154];

\bibitem{Simonov:2004ar}
  Yu.~A.~Simonov and J.~A.~Tjon,
  Phys.\ Rev.\  D {\bf 70}, 114013 (2004)
  [arXiv:hep-ph/0409361].

\bibitem{Eichten:2004uh}
  E.~J.~Eichten, K.~Lane and C.~Quigg,
  Phys.\ Rev.\  D {\bf 69}, 094019 (2004)
  [arXiv:hep-ph/0401210].


\bibitem{Kalashnikova:2005ui}
  Yu.~S.~Kalashnikova,
  Phys.\ Rev.\  D {\bf 72}, 034010 (2005)
  [arXiv:hep-ph/0506270].


\bibitem{Pennington:2007xr}
  M.~R.~Pennington and D.~J.~Wilson,
  Phys.\ Rev.\  D {\bf 76}, 077502 (2007)
  [arXiv:0704.3384 [hep-ph]].

\bibitem{Barnes:2007xu}
  T.~Barnes and E.~S.~Swanson,
  Phys.\ Rev.\  C {\bf 77}, 055206 (2008)
  [arXiv:0711.2080 [hep-ph]].


\bibitem{Danilkin:2009hr}
  I.~V.~Danilkin and Yu.~A.~Simonov,
  Phys.\ Rev.\  D {\bf 81}, 074027 (2010)
  [arXiv:0907.1088 [hep-ph]]; I.~V.~Danilkin and Yu.~A.~Simonov,
  Phys.\ Rev.\ Lett.\  {\bf 105}, 102002 (2010)
  [arXiv:1006.0211 [hep-ph]].

\bibitem{Zhang:2009bv}
  O.~Zhang, C.~Meng and H.~Q.~Zheng,
  Phys.\ Lett.\  B {\bf 680}, 453 (2009)
  [arXiv:0901.1553 [hep-ph]].


\bibitem{Li:2009ad}
  B.~Q.~Li, C.~Meng and K.~T.~Chao,
  Phys.\ Rev.\  D {\bf 80}, 014012 (2009)
  [arXiv:0904.4068 [hep-ph]].


\bibitem{Ortega:2010qq}
  P.~G.~Ortega, J.~Segovia, D.~R.~Entem and F.~Fernandez,
  Phys.\ Rev.\  D {\bf 81}, 054023 (2010)
  [arXiv:1001.3948 [hep-ph]].


\bibitem{Jacob:1959at}
  M.~Jacob and G.~C.~Wick,
  Annals Phys.\  {\bf 7} (1959) 404
  [Annals Phys.\  {\bf 281} (2000) 774].

\bibitem{Micu:1969a}
L.Micu, Nucl.Phys.B10, 521 (1969).

\bibitem{orsay}A.~Le Yaouanc, L.~Oliver, O.~Pene and J.~C.~Raynal, Phys.\ Rev.\ D {\bf 8} (1973)
2223; ibid.,Phys.\ Rev.\ D {\bf 9}, 1415 (1974); Phys.\ Rev.\ D {\bf
11}, 680 (1975); Phys.\ Rev.\ D {\bf 11}, 1272 (1975); Phys.\ Lett.\
B {\bf 71}, 397(1977).


\bibitem{Geiger:1994kr}
  P.~Geiger and E.~S.~Swanson,
  Phys.\ Rev.\  D {\bf 50}, 6855 (1994)
  [arXiv:hep-ph/9405238].

\bibitem{Ackleh:1996yt}
  E.~S.~Ackleh, T.~Barnes and E.~S.~Swanson,
  Phys.\ Rev.\  D {\bf 54}, 6811 (1996)
  [arXiv:hep-ph/9604355].

\bibitem{Barnes:1996ff}
  T.~Barnes, F.~E.~Close, P.~R.~Page and E.~S.~Swanson,
  Phys.\ Rev.\  D {\bf 55}, 4157 (1997)
  [arXiv:hep-ph/9609339].

\bibitem{Barnes:2002mu}
  T.~Barnes, N.~Black and P.~R.~Page,
  Phys.\ Rev.\  D {\bf 68}, 054014 (2003)
  [arXiv:nucl-th/0208072].


\bibitem{Liu:2010a}
  J.~F.~Liu, G.~J.~Ding and M.~L.~Yan,
  Phys.\ Rev.\  D {\bf 82}, 074026 (2010),[arXiv:1008.0246].


\bibitem{pdg} K.Nakamura et al. (Particle Data Group), J.P.G 37, 075021 (2010).

\bibitem{Godfrey:1985xj}
  S.~Godfrey and N.~Isgur,
  Phys.\ Rev.\ D {\bf 32}, 189 (1985).


\bibitem{Karliner:2008rc}
  M.~Karliner and H.~J.~Lipkin,
  arXiv:0802.0649 [hep-ph].


\bibitem{Ali:2010pq}
  A.~Ali, C.~Hambrock and S.~Mishima,
  Phys.\ Rev.\ Lett.\  {\bf 106}, 092002 (2011)
  [arXiv:1011.4856 [hep-ph]].

\bibitem{Hou:2006it}
  W.~S.~Hou,
  Phys.\ Rev.\  D {\bf 74}, 017504 (2006)
  [arXiv:hep-ph/0606016].

\bibitem{Meng:2008dd}
  C.~Meng and K.~T.~Chao,
  Phys.\ Rev.\  D {\bf 78}, 034022 (2008)
  [arXiv:0805.0143 [hep-ph]].

\bibitem{Zhou:1990ik}
  H.~-Y.~Zhou and Y.~-P.~Kuang,
  Phys.\ Rev.\ D {\bf 44}, 756 (1991).


\bibitem{Guo:2010ak}
  F.~-K.~Guo, C.~Hanhart, G.~Li, U.~-G.~Meissner and Q.~Zhao,
  Phys.\ Rev.\ D {\bf 83}, 034013 (2011)
  [arXiv:1008.3632 [hep-ph]].


\bibitem{Badalian:2012jz}
  A.~M.~Badalian, V.~D.~Orlovsky, Y.~.A.~Simonov and B.~L.~G.~Bakker,
  arXiv:1202.4882 [hep-ph].



\end{thebibliography}
\end{document}